\documentstyle[prb,aps,preprint]{revtex}
\newcommand{\bacugeo}{Ba$_2$CuGe$_2$O$_7$ }

\begin{document}

\draft

\title{Magnetic anisotropy and low-energy spin waves in the
Dzyaloshinskii-Moriya spiral magnet \bacugeo}

\author{A. Zheludev \and S. Maslov \and G. Shirane}

\address{Brookhaven National  Laboratory,
Upton, NY 11973-5000, USA.}

\author{I. Tsukada \and T. Masuda \and K. Uchinokura}

\address{Department of Applied Physics, The University of Tokyo,\\
7-3-1 Hongo, Bunkyo-ku, Tokyo 113-8656, Japan.}

\author{I. Zaliznyak,\footnote{Department of Physics and Astronomy
Johns Hopkins University, MD 21218 USA,\\ and P.~L.~Kapitza Institute
for Physical Problems, Moscow, Russia.} \and R. Erwin}
\address{NIST Center for Neutron Research, National Institute of Standards and
Technology, MD 20899}

\author{L. P. Regnault}
\address{DRFMC/SPSMS/MDN, CENG, 17 rue des Martyrs,\\ 38054 Grenoble Cedex, France}

\date{\today}

\maketitle

\begin{abstract}
New neutron diffraction and inelastic scattering experiments are used to
investigate in detail the field dependence of the magnetic structure and
low-energy spin wave spectrum of the Dzyaloshinskii-Moriya helimagnet
Ba$_2$CuGe$_2$O$_7$. The results suggest that the previously proposed
model for the magnetism of this compound (an ideal sinusoidal spin
spiral, stabilized by isotropic exchange and Dzyaloshinskii-Moriya
interactions) needs to be refined. Both new and previously published
data can be quantitatively explained by taking into account the
Kaplan-Shekhtman-Entin-Wohlman-Aharony (KSEA) term, a special magnetic
anisotropy term that was predicted to always accompany
Dzyaloshinskii-Moriya interactions in insulators.
\end{abstract}
\pacs{75.30.Et,75.10.Hk,75.30.Ds,75.30.Gw}

\section{introduction}

The recently discovered spiral magnet Ba$_2$CuGe$_2$O$_7$ is one of many
materials known to have incommensurate magnetic structures.\cite
{Zheludev96BACUGEO,Zheludev97BACUGEO} A fortunate combination of properties
however make Ba$_2$CuGe$_2$O$_7$ a particularly useful model system for
both experimental and theoretical studies of incommensurate magnetism: 1)
unlike the extensively studied rare-earth compounds,\cite{Jensen-Mackintosh}
Ba$_2$CuGe$_2$O$_7$ is an insulator, and thus can be conveniently described
in terms of localized spins; 2) Helimagnetism in Ba$_2$CuGe$_2$O$_7$ is
caused by the somewhat exotic Dzyaloshinskii-Moriya off-diagonal exchange
interactions\cite{Dzyaloshinskii57,Moriya60} that involve only
nearest-neighbor spins. This is in contrast with such well-known systems as
MnO$_{2}$ (Ref.~\onlinecite{Yoshimori59}) and NiBr$_{2}$
(Ref.~\onlinecite{Day80,Adam80}), where the magnetic incommensurability
results from a competition between exchange interactions for different
neighbor pairs (geometric frustration). 3) Compared to such classic
Dzyaloshinskii-Moriya helimagnets as MnSi (Ref.~
\onlinecite{Ishikawa76,Ishikawa84}) and FeGe (Ref.~\onlinecite{Lebech89}),
Ba$_2$CuGe$_2$O$_7$ has a rather low (tetragonal) crystal symmetry. The
result is a much richer field-temperature phase diagram. In particular,
in Ba$_2$CuGe$_2$O$_7$ a magnetic field applied along the unique
tetragonal axis induces a peculiar Dzyaloshinskii-type\cite{Dzya64}
incommensurate-to-commensurate
transition.\cite{ZM97BACUGEO-L,ZM98BACUGEO} Applying a magnetic field in
the tetragonal plane does not change the length of the magnetic
propagation vector, but leads to its re-orientation.\cite
{ZM97BACUGEO-B} 4) The spin arrangement in Ba$_2$CuGe$_2$O$_7$ is a
perfect square lattice. This fact allowed us previously to describe the
static properties of this remarkable system using a simple and elegant
macroscopic free energy
functional.\cite{ZM97BACUGEO-L,ZM98BACUGEO,ZM97BACUGEO-B} 5) Last but
not least, the scale of energies and wave numbers that characterize
magnetic interactions in Ba$_2$CuGe$_2$O$_7$ are very convenient for
neutron scattering measurements. Magnetic fields in which the most
interesting magnetic phase transformations occur are also readily
accessible using standard equipment.

As described in detail elsewhere (Refs.~
\onlinecite{Zheludev96BACUGEO,ZM98BACUGEO}),
the principal feature of Ba$_{2} $CuGe$_{2}$O$_{7}$ is a square-lattice
arrangement of Cu$^{2+}$ ions in the $(a,b)$ plane of the tetragonal
non-centric crystal structure (space group $P\overline{4}2_{1}m$,
$a=8.466$\AA ,$~c=5.445$\AA). Nearest-neighbor in-plane
antiferromagnetic exchange coupling [along the $ (1,1,0)$ direction] is
by far the strongest magnetic interaction in the system ($J\approx
0.96$~meV per bond\cite{oldJ}). The interaction between Cu-spins from
adjacent planes is much weaker and ferromagnetic ($J_{\bot }\approx
-0.026$ ~meV per bond). The magnetic structure can be described as an
almost-antiferromagnetic spiral (Fig.~1, insert), with spins confined in
the $(1,\overline{1} ,0)$ plane and the magnetic propagation vector
$(1+\zeta ,\zeta ,0),\zeta
\approx 0.0273$, $(1,0,0)$ being the N\'{e}el point. It was previously
demonstrated that the helimagnetic state is stabilized by nearest-neighbor
Dzyaloshinskii-Moriya interactions that for two interacting spins
$\bbox{S}_{1}$ and $\bbox{S}_{2}$ can be written as $(\bbox{S}_{1}\times
\bbox{S}_{2})\cdot \bbox{D}^{(1,2)}$. For the Cu-Cu bond along the $(1,1,0)$
direction ($x$-axis) the Dzyaloshinskii vector $\bbox{D}$ is pointing along
$ (1,\overline{1},0)$ ($y$-axis), inducing a relative rotation of the
interacting spins in the $(x,z)$ plane [the $z$ axis is chosen along the
$c$
-axis of the crystal]. The rotation angle $\phi $ (relative to a perfect
antiparallel alignment) is related to the magnetic propagation vector by $
\phi =2\pi \zeta \approx 0.172 $. Obviously, two types of domains, with
equivalent propagation vectors $(1+\zeta ,\zeta ,0)$ and $(1+\zeta ,-\zeta
,0)$ will always be present in a macroscopic sample.

By now, a large amount of experimental and theoretical work has been
done on Ba$_2$CuGe$_2$O$_7$, mainly dealing with the phase transitions
and static magnetic properties. Some important issues remain unresolved
however. For example, it was predicted that applying a magnetic field
along the unique axis should give rise to a distortion of the ideal
spiral structure.\cite {ZM97BACUGEO-L,ZM98BACUGEO} This so-called
soliton phase is characterized by the appearance of higher-order
magnetic Bragg harmonics. To date these additional Bargg reflections
have not been observed directly in an experiment. As far as the spin
dynamics is concerned, only the near-zone-boundary spin wave dispersion
relations were studied. For the physics of the incommensurate state, it
is the the low-energy, small-$Q$ spin excitations that are most
relevant. In the present work we continue our studies of
Ba$_2$CuGe$_2$O$_7$, investigating the field-dependence of higher-order
magnetic Bragg peaks and the low-energy spin-wave spectrum in both the
incommensurate and commensurate states. We find that even in the absence
of an external magnetic field the spiral structure is distorted by the
presence of previously disregared magnetic anisotropy. Our new results
are consistent with the theoretical predictions of
Kaplan,\cite{Kaplan83} Shekhtman, Entin-Wohlman and
Aharony,\cite{Shekhtman92,Shekhtman93} who demonstrated that a generic
anisotropy term must inevitably accompany Dzyaloshinskii-Moriya
interactions. The new understanding of the physics of
Ba$_2$CuGe$_2$O$_7$ enables us to refine our interpretation of
previously obtained experimental data. A brief report on some of our new
results is published elsewhere \cite{prl}.

\section{Experimental}

Neutron diffraction and inelastic neutron scattering measurements were
performed in two series of experiments, on the IN-14 3-axis spectrometer
at the Institut Laue Langevein (ILL) in Grenoble, and the SPINS
spectrometer at the Cold Neutron Research Facility at the National
Institute of Standards and Technology (NIST). Single-crystal samples of
Ba$_2$CuGe$_2$O$_7$ rarely survive more than one cooling to low
temperatures. Two different crystals, prepared by the floating-zone
method, were used in the two experimental runs. Sample A, a cylindrical
single crystal of dimensions $4\times 4\times 20$~mm$^3$ was used in
experiments at IN-14, but spontaneously disintegrated during subsequent
storage. Sample B was used in the second experiment on SPINS and was
approximately $6\times 6\times 50$~mm$^3$. Crystal mosaic was around
0.35$^{\circ}$ FWHM for sample A and 1.2$^{\circ}$ FWHM for sample B, as
measured in the $(a,b)$ crystallographic plane. The mosaic spread in the
perpendicular direction was measured for sample B and found to be around
2$
^{\circ}$ FWHM. The samples were mounted on the spectrometers with their $c$
-axes vertical, making $(h,k,0)$ wave vectors accesible for measurements. In
both experiments the magnetic field was produced by standard split-coil
superconducting magnets. The alignment of the $c$ axis of the crystal
with the direction of the magnetic field, previously shown to be crucial
for high-field measurements,\cite{ZM98BACUGEO} was around $1.4^{\circ}$
in both runs, as measured at low temperatures. The measurements were
performed in the field range 0--2.5~T. The sample environment was a
pumped-$^4$He cryostat for the ILL experiment and a cryopump-driven
$^3$He cryostat at NIST. The data were collected at temperatures in the
range 0.35--5~K. As observed previously, cooling the sample through
$T_{N}$ in an $H\approx1$~T magnetic field always resulted in a
single-domain magnetic structure.

The spin wave dispersion was measured in constant-$Q$ scans in the range of
energy transfers 0--0.8~meV. Neutrons of 3.5~meV or 2.5~meV fixed incident
energy were used in most cases. Alternatively, a 3.5~meV fixed final energy
setup was exploited. A Be filter was positioned in front of the sample to
eliminate higher-order beam contamination.
$40^{\prime}-S-40^{\prime}-A-40^{
\prime}$ collimations were utilized in both runs. The typical energy
resolution with 3.5~meV incident energy neutrons was 0.075~meV FWHM, as
determined from measurements of incoherent scattering from the sample.

\section{results}

\subsection{Higher-order Bragg reflections}

In previous studies the only magnetic elastic peaks observed in Ba$_2$CuGe$
_2 $O$_7$ were those corresponding to an ideal sinusoidal spiral structure
with propagation vectors $(1\pm\zeta,\pm\zeta,0)$. These reflections,
whose intensity seems to account for almost 100\% of the expected
magnetic diffraction intensity appear below $T_{N}\approx 3.2$~K around
antiferromagnetic zone-centers $(h,k,0)$, $h$, $k$-integer, $h+k$-odd.
In the present study, careful elastic scans along the $(1+x,x,0)$ line
in reciprocal space revealed the presence of additional extremely weak
peaks at $(1\pm 3\zeta,\pm 3\zeta,0)$, as shown in Fig.~\ref{diff}.
These peaks are clearly of magnetic origin, as the temperature
dependence of their intensity is similar to that of the principal
magnetic reflections at $ (1\pm\zeta,\pm\zeta,0)$. The additional
3rd-order peak was observed in all magnetic fields in the range
0--1.9~T, and always appears at $(1\pm 3\zeta,\pm 3\zeta,0)$, where
$\zeta$, defined by the position of the principal magnetic Bragg peak,
is itself field-dependent.\cite{ZM97BACUGEO-L,ZM98BACUGEO} For $
0<H<1.7$~T the measured field-dependence of $\zeta$ is in total
agreement with previous studies. For the purpose of convenience we shall
define $
\bbox{Q}_{\pi,\pi}=(1,0,0)$ (antiferromagnetic zone-center), and $\bbox{q}_0
=(\zeta,\zeta,0)$. In this notation the 1st and 3rd order magnetic
reflections correspond to momentum transfers
$\bbox{Q}_{\pi,\pi}\pm\bbox{q}_0$ and $\bbox{Q}_{\pi,\pi}\pm\bbox{3q}_0$,
respectively.

As observed previously, at $H=H_{1}\approx 1.7$~T (at $T=0.35$~K) the system
goes through a magnetic transition to a new phase that is characterized by
the appearance of a new peak at the commensurate $(1,0,0)$ reciprocal-space
position. As discussed previously, this new phase may or may not be a result
of the slight misalignment of the magnetic field relative to the $c$-axis of
the crystal. In the present work we did not investigate this
``intermediate'' phase in detail, performing most measurements in the field
ranges $0<H<H_{1}$ and $H>H_{c}\approx 2.2$~T, where $H_{c}$ is the magnetic
field at which the structure becomes commensurate.\cite
{ZM97BACUGEO-L,ZM98BACUGEO}

In scans along the $(1,1,0)$ direction, shown in Fig.~\ref{diff}, the
widths of both 1st and 3rd-order peaks are resolution-limited. This is
not the case for transverse scans along $(1,\overline{1},0)$, where the
1st, and especially the 3rd harmonic are visibly broader than the
experimental resolution (Fig.~\ref{diff2}). The observed peak width
pattern is consistent with both peaks having a zero longitudinal and a
20$^{\circ}$ transverse intrinsic widths {\it as seen from the
$\bbox{Q}_{\pi,\pi}$ reciprocal-space point}. The transverse intrinsic
$Q$-width of the 3rd harmonic is thus three times as large as that of
the first harmonic. This result does not appear to depend on the applied
magnetic field or the history of the sample. The large observed
transverse width is likely to be related to the previously established
fact that the spiral structures propagating in any direction in the
$(001)$ plane have almost identical energies. The $(1,1,0)$ direction is
only slightly energetically preferable.\cite{ZM97BACUGEO-B} Due to
pinning or even thermal fluctuations, in a macroscopic sample an entire
ensemble of spiral structures with propagation directions fanning out
around $(1,1,0)$ will therefore be realized, producing substantial
transverse peak widths.

The field dependence of the integrated peak intensities was measured in
both field-cooling and zero-field cooling experiments. Consistent
results were obtained in both types of measurement, and no signs of
hysteresis were observed. In the ILL experiment the propagation
direction of the spiral, always along $(1,1,0)$ at $H=0$, was found to
deviate by as much as several degrees from this direction in higher
fields. This effect is clearly due to a slight misalignment of the
magnetic field relative to the $c$-axis, and the possibility to almost
freely rotate the magnetic propagation vector in the $(a,b)$
crystallographic plane.\cite{ZM97BACUGEO-B} In the NIST experiment such
a deviation was not observed, thanks to a slightly different and more
``fortunate'' setting of the sample. The field dependence of the peak
intensities was therefore measured in this second experimental run, but,
just in case, at each field, both the 1st and 3rd-order peaks were
centered in a series of transverse and longitudinal scans. The measured
integrated intensity of the 1st and third-order reflection, as well as
that of the commensurate peak at $(1,0,0)$, are plotted against magnetic
field applied along the $c$-axis in Fig.~\ref{diffresult}. The total
intensity of all three features is field-independent within experimental
error.

As seen in Fig.~\ref{diffresult}, the intensity of the $(1+ \zeta,
\zeta,0)$ magnetic reflection is almost field-independent in the range
$ 0<H<H_{1}$. This appears to be in contradiction with previous
measurements (Ref.~\onlinecite{ZM98BACUGEO}, Fig.~3d), where a gradual
decrease of the intensity of the 1st harmonic was observed with
increasing magnetic field. However, we now know what was wrong with
these previous measurements: the possibility of the propagation vector
deviating from the $(1,1,0)$ direction was not taken into account. In a
slightly misaligned sample the field-induced drift of the magnetic
reflections away from the line of the elastic scan was incorrectly
interpreted as a decrease of peak intensity. Note that in the present
study the centering of the peaks at each field ensures that this
problem, even if present, does not influence the measurements.

\subsection{Spin waves}

All inelastic measurements were done in the vicinity of the $(1,0,0)$
antiferromagnetic zone-center ($\bbox{Q}\approx \bbox{Q}_{\pi,\pi}$).
The spin wave dispersion was measured along the
$(1+\epsilon,\epsilon,0)$ direction ($x$-axis). In most cases the sample
was field-cooled to eliminate the need to deal with inelastic signal
coming from the two magnetic domains. All scans at $\bbox{Q}=\bbox{
Q}_{\pi,\pi}$ were repeated using zero-field-cooling to ensure that no
hysteresis effects influence the measurements. The most important
limiting factor in these inelastic studies is the presence of incoherent
scattering and magnetic Bragg tails, centered at zero energy transfer.
This undesirable contaminations is absent for energy transfers
$(\hbar\omega)\gtrsim 0.15$ ~meV, where reliable data could be
collected. Typical constant-$Q$ scans obtained in the ILL and NIST
experiments are shown in Fig.~\ref{exdata}. Peak positions were
determined by fitting Gaussian profiles to the data. All the inelastic
peaks studied were found to be resolution-limited. The focusing
conditions are considerably more favorable at $\bbox{Q}
=(1+\epsilon,\epsilon,0)$, $\epsilon>0$, where most of the measurements were
performed.

\subsubsection{Zero field}

The dispersion relation measured in zero applied field is plotted in
symbols in Fig.~\ref{disp0}. One clearly sees three distinct branches of
the spectrum. These we shall label by the wave vectors to which they
extrapolate at zero energy transfer: $\bbox{Q}_{\pi ,\pi }\pm
\bbox{q}_{0}$ and $\bbox{Q}
_{\pi ,\pi }$, correspondingly. An obvious and very interesting feature is
the ``repulsion'' between the $\bbox{Q}_{\pi ,\pi }\pm \bbox{q}_{0}$
branches at their point of intersection $\bbox{Q}=\bbox{Q}_{\pi ,\pi }$.
Its magnitude is given by the splitting $2\delta _{\pi ,\pi }\approx
0.12(1)$~meV. This effect again manifests itself at $\bbox{Q}_{\pi ,\pi
}+2\bbox{q}_{0}$, where it is seen as a discontinuity in the
$\bbox{Q}_{\pi ,\pi }\pm \bbox{q}_{0}$ branch. Simple empirical fits to
the data (not shown) allow us to estimate the spin wave velocity
$c_{0}\approx 5.21(3)$~meV\AA .\ This value is in reasonable agreement
with the estimate $c_{0}={Ja}/\sqrt{2}\approx 5.75$~meV
\AA , obtained using the classical formula (3) in Ref.~\onlinecite{Zheludev96BACUGEO} and
the exchange constant $J\approx 0.96$~meV, previously determined from
measuring the spin wave bandwidth.\cite{} The main characteristic of the
$
\bbox{Q}_{\pi ,\pi }$ branch is the energy gap $\Delta _{\pi ,\pi }\approx
0.18(1)$~meV at the antiferromagnetic zone-center $\bbox{Q}_{\pi ,\pi }$.

\subsubsection{\label{incom}Field dependence in the incommensurate phase ($H<H_{1}$)}

In Figure ~\ref{disp1} we show the spin wave dispersion measured in
Ba$_2$ CuGe$_2$O$_7$ in a $H=1$~T magnetic field applied along the $c$
axis of the crystal. In this case the incommensurability parameter
$\zeta(H=1~{\rm T})=0.0252(5)$. The $\bbox{Q}_{\pi,\pi}\pm\bbox{q}_0$
dispersion curves are very similar to those measured in zero field and
appear to be adequately described by the same spin wave velocity and
splitting parameter $2\delta_{\pi,\pi}$. Compared to the zero field case
however, at $H=1$~T the central $\bbox{Q}
_{\pi,\pi}$ branch is visibly flattened at its minimum. The gap $
\Delta_{\pi,\pi}$ in this mode is equal to $\approx 0.24$~meV. Comparing
this to $\Delta_{\pi,\pi}=0.18$~meV at $H=0$, we find that, to a good
approximation:
\begin{equation}
\Delta_{\pi,\pi}(H)^2=\Delta_{\pi,\pi}^2+(2 g_{c} S \mu_{B} H)^2,
\label{edisp3}
\end{equation}
where $g_{c}=2.474$ is the $c$-axis diagonal component of the gyromagnetic
ratio for Cu$^{2+}$ in Ba$_2$CuGe$_2$O$_7$ (Ref.~\onlinecite{Sasago-ESR}),
$
S=1/2$ is the spin of Cu$^{2+}$ ions and $\mu_{B}$ is the Bohr magneton. At
$ H=1$~T the measured dispersion curve for the $\bbox{Q}_{\pi,\pi}$ branch
has a new feature, namely a discontinuity at
$\bbox{Q}_{\pi,\pi}+\bbox{q}_0$. This splitting, that we shall denote as
$2\delta_{\bbox{q}_0}$, is roughly $ 0.05$~meV.

In a magnetic field $H=1.5$~T$\simeq H_{c1}$
($\zeta=0.0232$) the spectrum becomes
substantially more complex (Fig.~\ref{disp15}). The two $\bbox{Q}
_{\pi,\pi}\pm\bbox{q}_0$ modes remain essentially unchanged. The $\bbox{Q}
_{\pi,\pi}$-gap in the central branch is $\Delta_{\bbox{Q}
_{\pi,\pi}}(H)\approx 0.28$~meV, which is consistent with Eq.~\ref{edisp3}.
The discontinuity at $\bbox{Q}_{\pi,\pi}+\bbox{q}_0$ is clearly visible in
$ H=1.5$~T data: $2\delta_{\bbox{q}_0}\approx 0.11$~meV. A new feature of
the spectrum that is not visible in lower applied fields is the presence of
a new excitation branch that at $\bbox{Q}_{\pi,\pi}$ is seen at at $(\hbar
\omega)\approx 0.45$~meV. The shortage of beam time prevented us from
following this branch to lower energy transfers (where its intensity should
increase) at wave vectors where it would appear focusing: all measurements
were done at $\bbox{Q}=(1+\epsilon,\epsilon,0)$, $\epsilon>0$. The limited
data that we have at this stage is totally consistent with the new branch
being a replica of the $\bbox{Q}_{\pi,\pi}$ mode, but centered at $\bbox{Q}
_{\pi,\pi}\pm 2\bbox{q}_0$, as shown by the corresponding solid lines in
Fig.~\ref{disp15}.

\subsubsection{High field: commensurate phase.}

The dispersion relations measured at $H=2.5$~T, well above $H_{c}\approx
2.2$ ~T, are shown in Fig.~\ref{disp25}. As expected for the commensurate
state, only two branches are present. Two peculiarities are to be noted
here. First, the measured spin wave velocity $c_{0}=4.83(3)$~meV\AA\ is
significantly smaller than that seen at $H<H_{c}$. Second, the gap in the
higher-energy branch ($\approx 0.45$~meV) is too large to be accounted for
by the effect of magnetic field alone ($2g_{c}S\mu _{B}H=0.36$~meV). If for
this branch we can write:
\begin{equation}
(\hbar \omega )^{2}=\Delta _{{\rm c}}^{2}+(2g_{c}S\mu
_{B}H)^{2}+c_{0}^{2}q^{2},
\label{edisp4}
\end{equation}
for the ``additional'' gap in the commensurate phase we obtain $\Delta
_{{\rm c}}=0.28(1)$~meV (solid lines in Fig.~\ref{disp25}).

\section{Theory}

Most of the magnetic properties of Ba$_{2}$CuGe$_{2}$O$_{7}$ reported to
date appeared to be rather well described by a simple spin Hamiltonian that
included only nearest-neighbor Heisenberg antiferromagnetic exchange
interactions and the Dzyaloshinskii-Moriya cross-product terms \cite{}. For
reasons that will shortly become apparent we shall refer to this construct
as the ``DM-only'' model for Ba$_{2}$CuGe$_{2}$O$_{7}$. For a single
Cu-plane in Ba$_{2}$CuGe$_{2}$O$_{7}$ the Hamiltonian takes the form:
\begin{eqnarray}
{\cal H} &=&{\cal H^{({\rm H})}+{H}^{({\rm DM})}=}
\sum_{n,m} \left[ J\ \left( \bbox{S}_{n,m}\cdot \bbox{S}_{n+1,m}+\bbox{S}
_{n,m}\cdot \bbox{S}_{n,m+1}\right)\right. +\nonumber \\
&+&\left. D\ [(\bbox{S}_{n,m}\times \bbox{S}
_{n+1,m})_{y}+(\bbox{S}_{n,m}\times \bbox{S}_{n,m+1})_{x} ] \right]
\label{hold}
\end{eqnarray}
Here the indexes $n$ and $m$ enumerate the Cu$^{2+}$ spins along the $x$
and $y$ axes, respectively, $\bbox{S}_{n,m}$ are the site spin operators,
$J$ is Heisenberg exchange constant and $D$ is the norm of the
Dzyaloshinskii vector. Microscopically, the Heisenberg term
$\cal {H}^{({\rm H})}$ represents the Anderson superexchange
mechanism.\cite{Anderson59} It arises from virtual {\it non-spin-flop}
hopping of two electrons onto a non-occupied orbital, where they interact
via Pauli's exclusion principle. As shown by Moriya, \cite{Moriya60} the
cross-product term $\cal{H}^{({\rm DM})}$ originates from {\it spin-flop}
hopping, which is made possible by spin-orbit interactions.

The classical ground state of the DM-only model is an ideal sinusoidal
spin-spiral. The experimental observation of higher-order magnetic
reflections in zero magnetic field tells us that this model is not a fully
adequate description of Ba$_{2}$CuGe$_{2}$O$_{7}$: something is missing
from the Hamiltonian (\ref{hold}). To understand what is going on we first
note that for two spins $\bbox{S}_{1}$ and $\bbox{S}_{2}$, interacting via
isotropic exchange and the Dzyaloshinskii-Moriya term, the interaction
energy is minimized at $-\sqrt{ J^{2}+D^{2}} \ S^{2}$ when both spins
$\bbox{S}_{1}$ and $\bbox{S}_{2}$ are perpendicular to $\bbox{D}$, forming
the angle $\pi +\alpha $, where $\alpha
=\arctan\left(D/J\right) $. Therefore, the Dzyaloshinskii-Moriya
cross-product term ${\cal H}^{({\rm DM})}$ lifts the local $O(3)$ symmetry
of the Heisenberg Hamiltonian and creates an effective easy plane
anisotropy of strength $
\sqrt{J^{2}+D^{2}}-J\simeq D^{2}/2J$.

Relatively recently Kaplan\cite{Kaplan83} and Shekhtman, Entin-Wohlman, and
Aharony\cite{Shekhtman92} (KSEA) argued that in most realizations of
Moriya's superexchange mechanism this apparent easy-plane anisotropy is an
artifact of the omission of terms quadratic in $D$ in the expansion of
the true Hamiltonian of the system. If such terms are
properly included, the $O(3)$ symmetry of a single bond is restored by an
additional term $(\sqrt{J^{2}+D^{2}}-J)/D^{2}\ (\bbox{S}_{1}\cdot
\bbox{D})(\bbox{S}
_{2}\cdot \bbox{D})\simeq (1/2J)(\bbox{S}_{1}\cdot \bbox{D})(\bbox{S}_{2}\cdot
\bbox{D})$. Note that this additional interaction has the form of easy-axis
two-ion anisotropy and its strength is such that it {\it exactly}
compensates the easy-plane effect of the Dzyaloshinskii-Moriya
cross-product. With this term included, the ground state of the two-spin
Hamiltonian has full $O(3)$ symmetry. The energy of two interacting
spins pointing parallel and antiparallel to $\bbox{D}$, respectively, is
exactly equal to that of two spins perpendicular to $\bbox{D}$ and
forming the angle $\pi +\alpha $ between themselves. We shall refer to
this ``hidden symmetry'' term as the KSEA anisotropy term or KSEA
interaction. For a recent discussion of this subject see Ref.
\onlinecite{Yildirim95}.

To properly account for KSEA interactions in our model of
Ba$_{2}$CuGe$_{2}$O$_{7}$, where $D \ll J$,
the spin Hamiltonian can be rewritten as
follows:
\begin{eqnarray}
{\cal H} &=&{\cal {H}^{({\rm H})}+{H}^{({\rm DM})}+{H}^{({\rm KSEA})}=}
\nonumber
\\
&&\sum_{n,m}\left[ J\ (\bbox{S}_{n,m}\cdot
\bbox{S}_{n+1,m}+\bbox{S}_{n,m}\cdot \bbox{S}_{n,m+1})+D\ \left(
(\bbox{S}_{n,m}\times \bbox{S}_{n+1,m})_{y}+
(\bbox{S}_{n,m}\times \bbox{S}_{n,m+1})_{x}\right)+\right.  \nonumber \\
&&+\left. {\frac{D^{2}}{2J}}
(S_{n,m}^{y}S_{n+1,m}^{y}+S_{n,m}^{x}S_{n,m+1}^{x})\right]  \label{ham}
\end{eqnarray}
Can this Hamiltonian (the ``DM+KSEA'' model) account for both new and
previously published experimental data on Ba$_{2}$CuGe$_{2}$O$_{7}$? In
the following sections we shall systematically investigate the effect of
the KSEA term on static and dynamic properties of a DM helimagnet, and
show that indeed it can.

\subsection{Static properties}

\subsubsection{\label{freee}Free energy in the continuous limit}
As the period of the spiral structure in Ba$_{2}$CuGe$_{2}$O$_{7}$ is rather
long ($\approx 36$ lattice spacings), we
can safely use the continuous
approximation to describe it.\cite{ZM98BACUGEO,ZM97BACUGEO-B}
In this framework the
magnetic free energy is expanded as a functional of a slowly rotating
unitary vector field $\bbox{n}(\bbox{r})$. At each point in space
$\bbox{n}(\bbox{r})$ is chosen along the local staggered magnetization. The
Hamiltonian (\ref {hold}) then gives rise to the following free energy
functional:
\begin{eqnarray}
F^{{\rm (DM)}} &=&\int dxdy\ \left[ {\frac{\rho _{s}}{2}}\left( (\partial
_{x}\bbox{n}-{\alpha \over \Lambda} \
\bbox{e}_{y}\times \bbox{n})^{2}+(\partial _{y}\bbox{n}
-{\alpha \over \Lambda}\ \bbox{e}_{x}\times \bbox{n})^{2}
\right) -\right.
\nonumber \\ &&\left. -{\frac{\alpha ^{2}}{2 \Lambda ^2}}
\rho _{s}n_{z}^{2}+{\frac{(\chi _{\perp }-\chi _{\Vert })(\bbox{H}\cdot
\bbox{n})^{2}}{2}}-{\frac{\chi _{\perp }H^{2}}{2}}
\right] .  \label{fold}
\end{eqnarray}
In this formula $\rho _{s}$ is the spin stiffness, that in the classical
model at $T=0$ is given by $\rho _{s}=S^{2}\sqrt{J^{2}+D^{2}}\approx
S^{2}J$, $\alpha $ as before is the equilibrium angle between two spins
defined as $\alpha =\arctan (D/J)$, $\Lambda$ is the nearest-neighbor
Cu-Cu distance, $\chi _{\Vert }$
and $\chi _{\bot }$ are the local longitudinal and transverse magnetic
susceptibilities, respectively. Their classical $T=0$ values are $\chi
_{\bot }=(g\mu _{B})^{2}/(4J\Lambda ^{2})$ and $\chi _{\Vert }=0$,
correspondingly. In Eq.~\ref{fold} we have included the Zeeman term that
represents interaction of the system with an external magnetic field
$\bbox{H}$.

The term $-\alpha ^{2} \rho _{s}n_{z}^{2}/2 \Lambda^2$ in Eq.~\ref{fold}
deserves some comment. It has the form of a magnetic easy-$z$-axis
anisotropy and represents the combined effect of the effective $(xz)$ and
$(yz)$ easy planes produced by DM interactions on the $y$- and $x$-bonds,
respectively. This term is {\it eliminated} by KSEA interactions that
modify Eq.~\ref{fold} as follows:
\begin{eqnarray}
F^{{\rm (DM+KSEA)}}= &=&\int dxdy\ \left[ {\frac{\rho _{s}}{2}}\left( (\partial
_{x}\bbox{n}-{\alpha \over \Lambda} \
\bbox{e}_{y}\times \bbox{n})^{2}+(\partial _{y}\bbox{n}
-{\alpha \over \Lambda}\ \bbox{e}_{x}\times \bbox{n})^{2}
\right) +\right.
\nonumber \\ &&\left. +{\frac{(\chi _{\perp }-\chi _{\Vert })(\bbox{H}\cdot
\bbox{n})^{2}}{2}}-{\frac{\chi _{\perp }H^{2}}{2}}
\right] . \label{free}
\end{eqnarray}
This equation is in agreement with Eq.(3) in
Ref.~\onlinecite{ZM97BACUGEO-B}. Comparing Eqs.~\ref{hold} and
\ref{free} one concludes that  {\it in the continuous limit}
for the square-lattice spin arrangement found in Ba$_{2}$CuGe$_{2}$O$_{7}$,
KSEA interactions (two sets of easy axes, for $x$ and $y$-bonds,
respectively) are indistinguishable from an overall easy-$(xy)$-plane
anisotropy of relative strength $\delta=\alpha ^{2}/2$.

In this work we are mostly concerned with the effect of a magnetic field
applied along the $[001]$ crystallographic direction, i.e., along the
$z$-axis. Under these conditions the propagation direction of the spin
spiral in Ba$_{2}$CuGe$_{2}$O$_{7}$ is either along the $x$ or $y$ axis
(two domain types are possible). Moreover, as we shall prove rigorously
while discussing the spin waves in the system, the magnetic structure
remains {\it planar} despite the two types of Dzyaloshinskii vectors,
along the $x$ and $y$ axes (for the $y$ and $x$ bonds, respectively).
This fact allows us to write the components of vector
$\bbox{n}(\bbox{r})$ as $(\sin \theta (x),0,\cos \theta (x))$, where
$\theta (x)$ is the angle between local staggered moment $\bbox{n
}(\bbox{r})$ and the $z$ axis, for a helix propagating along the $x$
direction. The free energy can be then rewritten in terms of the $\theta
(x)$ as
\begin{equation}
F^{{\rm (DM+KSEA)}}=\int dx\ dy\ \left[ {\frac{\rho _{s}
(\partial_{x}\theta
-(\alpha/ \Lambda))^{2}}{2}}+({\alpha^2 \rho _{s} \over 2 \Lambda^2}
+{\frac{(\chi _{\perp }-\chi _{\Vert })H^{2}}{2
}})\cos ^{2}\theta -{\frac{\chi _{\perp }H^{2}}{2}}\right] .
\end{equation}
This is {\it exactly} Eq. (1) of Ref.~\onlinecite{ZM97BACUGEO-L} modified
to include the effects of an easy $(xy)$ plane anisotropy
$\rho_s \alpha ^{2} n_{z}^2/2 \Lambda^2=
\mbox{\rm const}-\rho_s \alpha ^{2} cos^2 \theta /2 \Lambda^2$,
coming from the KSEA interaction on
$y$ bonds.  As seen from this equation the sole effect of such anisotropy is
to renormalize the external field to
\begin{equation}
H_{{\rm eff}}(H)=\sqrt{H^{2}+ \alpha^2 \rho _{{\rm s}}/\Lambda^2
(\chi _{\perp }-\chi _{\Vert })}.
\label{heff}
\end{equation}

\subsubsection{Critical field and magnetic propagation vector}
One important consequence of what is said above is that all our previous
results, obtained in Refs.~\onlinecite{ZM97BACUGEO-L,ZM98BACUGEO}, can
be {\it recycled} by substituting $H_{{\rm eff}}(H)$ for $H$ in all
formulas. Our conclusions regarding the field-induced
commensurate-incommensurate Dzyaloshinskii transition in
Ba$_{2}$CuGe$_{2}$O$_{7}$ remain valid in the presence of KSEA
interaction. The KSEA interaction, however, modifies the value of the
critical field $H_{{\rm c}}$. Indeed, substituting $H_{{\rm eff}}(H)$
for $H$ in the Eq. 5 of Ref.~\onlinecite{ZM97BACUGEO-L} one gets
$\sqrt{H_{{\rm c}}^{2}+ \alpha^2/\Lambda^2 \rho _{{\rm
s}}/(\chi_{\bot}-\chi_{\|})}= (\pi
\alpha/2
\Lambda) \sqrt{\rho _{{\rm s}}
/(\chi _{\bot }-\chi _{\Vert })}$.
>From this we immediately obtain:
\begin{equation}
H_{{\rm c}}=\alpha {\frac{\sqrt{\pi ^{2}-4}}{2 \Lambda}}\sqrt{\frac{
\rho _{{\rm s}}}{\chi _{\bot }-\chi _{\Vert }}}\text{.}  \label{hc}
\end{equation}
We see that the KSEA term reduces the critical field by the universal
factor $\sqrt{1-4/\pi ^{2}}\simeq 0.771$.

In order to obtain the field dependence of the inverse period of the
structure $\zeta$ one has to rewrite  Eqs. (4,7) of
Ref.~\onlinecite{ZM97BACUGEO-L} as
\begin{eqnarray}
\frac{2 \pi \zeta (H)}{\alpha} &=&\frac{\pi ^{2}}{4E(\beta )K(\beta )}
\label{zeta1} \\
\frac{ H_{{\rm eff}}(H)}{H_{{\rm eff}}(H_{{\rm c}})}
&=&\frac{\beta }{E(\beta )}\text{.}
\label{zeta2}
\end{eqnarray}
Here $\beta $ is an implicit variable. In case when the deformation of
the spiral is weak ($(\alpha- 2\pi \zeta(H))/\alpha \ll 1$) one can
safely use the linearized formula:
\begin{equation}
\frac{2 \pi \zeta (H)}{\alpha} =1-{1 \over 32}
\left(\frac{\pi H_{{\rm eff}}(H)}{2 H_{{\rm eff}}(H_{{\rm c}})} \right)^4 +
\mbox{\rm higher order terms}.
\label{linear_zeta}
\end{equation}
From this formula one can derive by how much KSEA interactions deform
the spiral in zero external field. Let us define $\phi=2 \pi \zeta(0)$
as the average angle between spins in the spiral in zero field. This
parameter is easily accessible experimentally and equal to $\phi=2 \pi
\cdot 0.0273=0.172 \simeq 10^{o}$.

Recalling that
$H_{{\rm eff}}(0)/H_{{\rm eff}}(H_{{\rm c}})=2/\pi$ and plugging it in
Eq. \ref{linear_zeta} one gets $\phi/\alpha \simeq 1-1/32$ or
\begin{equation}
\alpha \equiv \arctan (D/J)=\frac{32}{31}\phi =\frac{32}{31}2\pi \zeta(0)
\text{.}  \label{alpha}
\end{equation}
The KSEA term thus increases the period of the
structure in zero field by roughly 3\%.

\subsubsection{\label{higher}Higher-order Bragg harmonics}
An important implication of Eq.~\ref{heff} is that even in {\it zero}
applied field the {\it effective} field is non-zero. The result is that the
spiral structure is distorted even in zero field and higher-order (odd)
magnetic Bragg peaks are present. To obtain a theoretical form for the
field dependence of the 3rd harmonic we can use Eqs. 17, 18
in Ref.~ \onlinecite{ZM98BACUGEO}. More practical than the resulting
expression is its linearized form, that applies in the limit
$(\alpha- 2\pi \zeta(H)) \ll  \alpha$ (weakly distorted spiral) :
\begin{equation}
{\frac{I_{3}}{I_{1}}}=
{1 \over 256}
\left(\frac{\pi H_{{\rm eff}}(H)}{2 H_{{\rm eff}}(H_{{\rm c}})}
\right)^4={1 \over 256}
\left[1+\left( {\frac{\pi ^{2}}{4}}- 1 \right)
\left( {\frac{H}{H_{{\rm c}}}}\right) ^{2}\right] ^{2}.
\label{harm}
\end{equation}
Here $I_{1}$ and $I_{3}$ are the intensities of the first and third
harmonic, respectively. One can see that for $H=0$ the third harmonic
is predicted to be smaller than the first one by a factor of
$1/256 \simeq 3.9 \times 10^{-3}$.

Comparing Eqs.~\ref{harm} and \ref{linear_zeta} one can see that
for weak distortions the intensity of the third harmonic is
proportional to the relative decrease of $\zeta(H)$:
\begin{equation}
{\frac{\zeta (H)}{\zeta (0)}}=1-{\frac{8(I_{3}(H)-I_{3}(0))}{I_{1}}}.
\label{harm2}
\end{equation}

\subsubsection{Comparison with experiment}

We now have to make sure that our results for the DM+KSEA model are
consistent with both the previously published
(Ref.~\onlinecite{ZM97BACUGEO-L,ZM98BACUGEO}) and new neutron
diffraction results on Ba$_{2}$CuGe$_{2}$O$_{7}$ . First, let us compare
the previously measured $\zeta (H)$ curve with the predictions of
Eqs.~\ref{zeta1},\ref{zeta2}. The experimental data for the
incommensurability parameter (Ref.~\onlinecite{ZM97BACUGEO-L}) is
plotted against $H^2$ in Fig.~\ref{maslov1}. The solid line is the
prediction of Eqs.~\ref{zeta1},\ref{zeta2}. The dashed line is the
theoretical curve previously obtained without including the KSEA term in
the Hamiltonian.\cite{ZM97BACUGEO-L,ZM98BACUGEO} In plotting both these
curves we have assumed the actual (measured) values for $H_{{\rm
c}}=2.15$~T and $\zeta(0)=0.0273$. Within experimental statistics it is
practically impossible to distinguish between the two theoretical
dependences and the data fits both of them reasonably well.

While it appears that the {\it shape} of the $\zeta (H)$ curve can not be
used to extract information on KSEA interactions, the actual numerical
value of $H_{{\rm c}}$ in the DM-only and DM+KSEA models is substantially
different. For the low-temperature limit in Ba$_{2}$CuGe$_{2}$O$_{7}$ we
can use the classical expressions $\rho
_{{\rm s}}=JS^{2}=0.24$~meV, $\chi _{\Vert }=0$ and
$\chi _{\perp }=(g_{c}\mu _{B})^{2}/8J \Lambda$, where $g_{c}=2.47$ is
the $c$-axis gyromagnetic ratio for Cu$^{2+}$ in
Ba$_{2}$CuGe$_{2}$O$_{7}$.\cite{Sasago-ESR} One can expect these
classical estimates to be rather accurate, as they rely on the {\it
effective} exchange constant $J$, that itself was determined from
fitting the {\it classical} spin wave dispersion relations to inelastic
neutron scattering data.\cite {Zheludev96BACUGEO} Substituting these
values into Eq.~\ref{hc} we immediately obtain $H_{{\rm c}}^{{\rm
(DM+KSEA)}}=2$~T. This is much closer to the experimental value $H_{{\rm
c}}=2.15$~T, than the estimate $H_{{\rm c}}^{{\rm (DM)}}=2.6$~T for the
DM-only model.

The new data for the field dependence of relative intensities of the first
and third Bragg harmonics becomes consistent with theory only if KSEA
interactions are properly taken into account. Indeed, the KSEA term is
necessary to reproduce the observed distortion of the spiral in zero
magnetic field. In Fig.~\ref{diffresult}(b) the solid lines are plotted
using Eq.\ref{harm} and $H_{{\rm c}}=2.15$~T. The dashed
lines are results for the intensity of the third Bragg harmonic obtained
previously for the DM-only model.\cite{ZM98BACUGEO} Clearly the DM+KSEA
model gives an excellent agreement with experiment (open symbols in
Fig.~\ref{diffresult}(b)), while the DM-only Hamiltonian fails entirely
to account for the available data.

In Fig. ~\ref{diffresult}(c) we check the validity of theoretical
prediction of Eq.\ref{harm2}, which is supposed to hold both with and
without KSEA terms. The excellent agreement of theory and
experiment confirms the validity of our picture of weakly deformed
almost sinusoidal spiral.

\subsection{\label{sd} Spin dynamics}
We now turn to calculating the classical spin wave spectrum in the DM+KSEA
model for Ba$_{2}$CuGe$_{2}$O$_{7}$. This task will be accomplished in
several separate steps. First, we shall derive the spectrum for a
square-lattice Heisenberg Hamiltonian, including DM interactions only for
the $x$-axis bonds. Second, we shall consider the effect of DM interactions
along the $y$
-axis bonds, showing that they do not disturb the planar spiral structure
and do not influence the dispersion relation along the $x$-direction. Next
we shall analyze the effect of adding the KSEA term, following the method
described in Ref.\onlinecite{Zaliznyak95,Zhitomirsky96}. While at this stage we
do not have results for spin wave dispersion in the DM+KSEA model in the
presence of an arbitrary external magnetic field, we shall consider the
case $H>H_{{\rm c}}$ and derive an expression for $\Delta_{{\rm c}}$
-- the anisotropy gap in the commensurate state.

\subsubsection{\label{simplest}
Dzyaloshinskii-Moriya interactions for $x$-axis bonds only.}

We start from a truncated version of the Hamiltonian (\ref{ham}):
\begin{equation}
{\cal H}^{(1)}=\sum_{n,m}[J\bbox{S}_{n,m}\cdot \bbox{S}_{n+1,m}+J\bbox{S}
_{n,m}\cdot \bbox{S}_{n,m+1}+D(\bbox{S}_{n,m}\times \bbox{S}_{n+1,m})_{y}]
\label{hamil1}
\end{equation}
It is easy to see that classically this Hamiltonian is minimized by a
perfect helicoid propagating along the $x$-axis with all spins lying in the
$(xz)$ plane:
\begin{eqnarray}
\langle {S_{n,m}^{z}}\rangle &=&(-1)^{n+m}\langle {S}\rangle \cos n\alpha ;
\nonumber \\
\langle {S_{n,m}^{x}}\rangle &=&(-1)^{n+m}\langle {S}\rangle \sin n\alpha .
\label{spin_conf}
\end{eqnarray}
The standard procedure to calculate the spin-wave spectrum is to rewrite
this Hamiltonian in terms of spin projections on the new rotating
coordinate system, where the direction of the equilibrium value of spin at
$(n,m)$ defines the {\it local} $z^{\prime }$ axis in such a way that $
\langle {S_{n,m}^{z^{\prime }}}\rangle =(-1)^{n+m}\langle {S}\rangle $. We
leave the $y$ coordinate unchanged, and select the new $x^{\prime }$
axis to be orthogonal to both $z^{\prime }$ and $y^{\prime }=y$.
Substituting $ S_{n,m}^{z}=S_{n,m}^{z^{\prime }}\cos n\alpha
-S_{n,m}^{x^{\prime }}\sin n\alpha $, $S_{n,m}^{x}=S_{n,m}^{x^{\prime
}}\cos n\alpha +S_{n,m}^{z^{\prime }}\sin n\alpha $, and
$S_{n,m}^{y}=S_{n,m}^{y^{\prime }}$ in the Hamiltonian (\ref{hamil1})
and using $\alpha =\arctan (D/J)$ we obtain:
\begin{equation}
{\cal H}^{(1)}=
\sum_{n,m}[\sqrt{J^{2}+D^{2}}(S_{n,m}^{z^{\prime}}S_{n+1,m}^{z^{\prime }}+
S_{n,m}^{x^{\prime }}S_{n+1,m}^{x^{\prime}})+JS_{n,m}^{y^{\prime
}}S_{n+1,m}^{y^{\prime }}+J\bbox{S}_{n,m}^{\prime }\cdot
\bbox{S}_{n,m+1}^{\prime }].
\end{equation}
In these coordinates the Hamiltonian is simply that of a square lattice AFM
with easy-plane exchange anisotropy on bonds along the $x$ direction. In
agreement with the discussion in Section~\ref{freee} the relative strength
of this anisotropy is given by:
\begin{equation}
\delta ={\frac{\sqrt{J^{2}+D^{2}}-J}{J}}\simeq {\frac{D^{2}}{2J^{2}}}\simeq {
\frac{\alpha ^{2}}{2}}.
\end{equation}

Applying the Holstein-Primakoff formalism we write the spin projection
operators as $S_{n,m}^{z^{\prime }}=(-1)^{n+m}(S-a_{n,m}^{\dagger
}a_{n,m})$, $S^{x^{\prime }}=(-1)^{n+m}
\ \sqrt{S/2}(a_{n,m}+a_{n,m}^{\dagger })$, $S^{y^{\prime
}}=i\sqrt{S/2}\
(a_{n,m}^{\dagger }-a_{n,m})$. From Eq. (\ref{hamil1}) it is then
straightforward to extract the quadratic part of the spin wave Hamiltonian:
\begin{eqnarray}
{\cal H}^{(1)} &=&JS\sum_{n,m}(4a_{n,m}^{\dagger
}a_{n,m}-a_{n,m}a_{n+1,m}-a_{n,m}a_{n,m+1}-a_{n,m}^{\dagger
}a_{n+1,m}^{\dagger }-a_{n,m}^{\dagger }a_{n,m+1}^{\dagger })- \nonumber \\
&-&\delta (a_{n,m}^{\dagger }+a_{n,m})(a_{n+1,m}^{\dagger
}+a_{n+1,m})/2+2\delta a_{n,m}^{\dagger }a_{n,m}.\label{az}
\end{eqnarray}
After performing Fourier and Bogolyubov transformations to diagonalize this
Hamiltonian, one readily obtains the spin wave spectrum:
\begin{equation}
{\cal E}(k_{x},k_{y})=JS\sqrt{[4+2\delta -\delta \cos k_{x}]^{2}-[(2+\delta
)\cos k_{x}+2\cos k_{y}]^{2}}.  \label{spectrum}
\end{equation}
This spectrum has one Goldstone
branch at $k_{x}=k_{y}=0$, that corresponds to the continuous symmetry of a
simultaneous rotation of all spins in the $(xz)$ plane.
At $k_x=k_y=\pi$ the spectrum has a finite gap $4JS \sqrt{\delta}=
2\sqrt{2} DS$ due to the easy $(xz)$ plane anisotropy coming from
the DM without KSEA correction.

Now we have to recall that in the above derivation the wave vectors $
k_{x},k_{y}$ correspond to a {\it rotating} system of coordinates. They are
thus distinct from the actual component of the scattering vector in a
neutron experiment. To get the proper spin wave spectrum one has to perform
a reverse coordinate transformation to the laboratory system:
\begin{eqnarray}
S_{n,m}^{z} &=&S_{n,m}^{z^{\prime }}\cos n\alpha -S_{n,m}^{x^{\prime }}\sin
n\alpha =  \nonumber \\
&=&(-1)^{n+m}[(S-a_{n,m}^{\dagger }a_{n,m})\cos n\alpha -\sqrt{S/2}\
(a_{n,m}+a_{n,m}^{\dagger })\sin n\alpha ];  \nonumber \\
S_{n,m}^{x} &=&S_{n,m}^{x^{\prime }}\cos n\alpha +S_{n,m}^{z^{\prime }}\sin
n\alpha =  \nonumber \\
&=&(-1)^{n+m}[(S-a_{n,m}^{\dagger }a_{n,m})\sin n\alpha +\sqrt{S/2}\
(a_{n,m}+a_{n,m}^{\dagger })\cos n\alpha ];  \nonumber \\
S_{n,m}^{y} &=&S_{n,m}^{y^{\prime }}=i\sqrt{S/2}\ (a_{n,m}^{\dagger }-a_{n,m}).
\label{unrot}
\end{eqnarray}
The $x$-axis dispersion of three spin wave branches in laboratory
system is shown in
\ref{tfig1}(a). The dynamic structure factor $S^{yy}(\bbox{Q},\omega )$ has
a single magnon peak at the energy given by ${\cal E}(Q_{x},Q_{y})$ (the
$\bbox{Q}_{\pi ,\pi }$-branch). The structure factors $S^{xx}(
\bbox{Q},\omega )$ and $S^{zz}(\bbox{Q},\omega )$ each contain two magnon
branches with dispersion relations given by ${\cal E}(Q_{x}+\pi +\alpha
,Q_{y}+\pi )$, and ${\cal E}(Q_{x}+\pi -\alpha ,Q_{y}+\pi )$ (the $\bbox{Q}
_{\pi ,\pi }\pm \bbox{q}_{0}$-branches). As expected, the zeroes of energy in
these two modes are precisely at the positions of magnetic Bragg peaks at
$\bbox{Q}_{\pi,\pi}\pm \bbox{q}_0$. A curious feature of this plot is that
all three branches are nearly degenerate at the AFM zone center.

\subsubsection{\label{DMY}Dzyaloshinskii-Moriya interactions along $y$-axis bonds}

Let us now consider Dzyaloshinskii-Moriya interactions for the bonds in
the $y$-direction. Their contribution to the spin Hamiltonian can be
written as:
\begin{eqnarray}
{\cal H}^{(2)}
&=&\sum_{n,m}D(S_{n,m}^{y}S_{n,m+1}^{z}-S_{n,m}^{z}S_{n,m+1}^{y})=
\sum_{n,m}DS_{n,m}^{y}(S_{n,m+1}^{z}-S_{n,m-1}^{z})=  \nonumber \\
&=&iDS\sum_{n,m}(-1)^{n+m}\sin n\alpha \ \lbrack
a_{n,m}a_{n,m+1}-a_{n,m}^{\dagger }a_{n,m+1}^{\dagger }]+
\text{{third order terms}}  \label{DM_y}
\end{eqnarray}
The absence of terms of the first order in $a_{n,m}$ and
$a_{n,m}^{\dagger }$ means that in the original (flat-spiral) spin
configuration the force acting on each spin produced by the added
Dzyaloshinskii-Moriya coupling on the $y$ bonds is equal to zero. Thus,
switching on the $y$-axis DM interactions {\it does not disturb} the
planar helimagnetic ground state of the Hamiltonian ${\cal H}^{(1)}$,
which therefore is also the ground state of ${\cal H}^{(2)}\equiv {\cal
H}^{({\rm H})}+{\cal H}^{({\rm DM})}$. This {\it a posteriori} verifies
our assumption that spins continue to lie in the $x-z$ plane in the
presence of Dzyaloshinskii-Moriya interactions on $y$-bonds, made in the
section (IV A 1).

New terms {\em quadratic} in $a_{n,m}$ and $a_{n,m}^{\dagger }$ are indeed
introduced by the Dzyaloshinskii-Moriya interactions on $y$-bonds,
and the spin wave spectrum is thus altered. After Fourier
transformation Eq. (\ref{DM_y}) becomes:
\begin{eqnarray}
{\cal H}^{(2)} &=&{\frac{iDS}{2}}\sum_{k_{x},k_{y}}\sin k_{y}\ [a^{\dagger
}(k_{x},k_{y})a^{\dagger }(-k_{x}+\pi +\alpha ,-k_{y}+\pi )-  \nonumber \\
&-&a^{\dagger }(k_{x},k_{y})a^{\dagger }(-k_{x}+\pi -\alpha ,-k_{y}+\pi )+
\nonumber \\
&+&a(k_{x},k_{y})a(-k_{x}+\pi -\alpha ,-k_{y}+\pi )-  \nonumber \\
&-&a(k_{x},k_{y})a(-k_{x}+\pi +\alpha ,-k_{y}+\pi )].  \label{DM_y_k}
\end{eqnarray}
The analysis of this term for general $k_{y}$ is rather complicated and
should be done by matrix diagonalization similar to that described in
the next subsection for calculating the effects of KSEA interactions.
Fortunately, for spin waves propagating along the $x$-axis ($k_x=0$ or
$k_x=\pi$) the contribution of ${\cal H}^{(2)}$ is exactly zero, thanks
to the $\sin k_{y}$ prefactor in Eq.~\ref{DM_y_k}. In other words, as
long as we are concerned with spin waves propagating along the (110)
direction in Ba$_{2}$CuGe$_{2}$O$_{7}$ we can totally disregard the
contribution of Dzyaloshinskii-Moriya interactions along the $y$-axis
bonds.

\subsubsection{Influence of KSEA interactions.}

Having understood the spectrum for the DM-only model, we can proceed to
include KSEA terms in our calculations. We first note that if our system
were {\it 1-dimensional}, the inclusion of the KSEA term would fully
restore $O(3)$ symmetry, making the commensurate and spiral phases
degenerate. In terms of spin waves this would signify a complete
softening of the $\bbox{Q}_{\pi}$ magnon branch at the AF zone-center
$\bbox{Q}_{\pi}$. As will be demonstrated below, in the case of a
2-dimensional spin arrangement in Ba$_{2}$CuGe$_{2}$O$_{7}$ the magnon
softening at $\bbox{Q}_{\pi,\pi}$ produced by KSEA interactions is
incomplete.

As discussed in Section~\ref{higher}, in the presence of the KSEA term
the ground state is a{\em\ distorted } flat spin-spiral. In this
situation the transition to a uniformly rotating coordinate system used
in Section~\ref{simplest} loses its usefulness. Instead, we must rotate
the coordinate system for spin quantization at each site in such a way,
that the $z$-axis follows the rotation of the spins in the distorted
helix:
\begin{eqnarray}
S_{n,m}^{z} &=&(-1)^{n+m}[(S-a_{n,m}^{\dagger }a_{n,m})\cos \theta _{n,m}-
\sqrt{S/2}\ (a_{n,m}+a_{n,m}^{\dagger })\sin \theta _{n,m}];  \nonumber \\
S_{n,m}^{x} &=&(-1)^{n+m}[(S-a_{n,m}^{\dagger }a_{n,m})\sin \theta _{n,m}+
\sqrt{S/2}\ (a_{n,m}+a_{n,m}^{\dagger })\cos \theta _{n,m}];  \nonumber \\
S_{n,m}^{y} &=&S_{n,m}^{y^{\prime }}=i\sqrt{S/2}(a_{n,m}^{\dagger }-a_{n,m}).
\label{gen_spin_conf}
\end{eqnarray}
Here $\theta _{n,m}$ denotes the angle between the local spin axis and $z$
-axis in the $x-z$ plane. The Hamiltonian ${\cal H}^{(1)}+{\cal H}^{(3)}$ \
(as explained above, ${\cal H}^{(2)}$ is not relevant to the dispersion along
the $x$-axis that we are interested in) is then rewritten as:
\begin{eqnarray}
{\cal H}^{(1)}+{\cal H}^{(3)} &=&J\sum_{n,m}\left[ -{\frac{\cos (\theta
_{n+1,m}-\theta _{n,m}-\alpha )}{\cos \alpha }}\left( (S-a_{n,m}^{\dagger
}a_{n,m})(S-a_{n+1,m}^{\dagger }a_{n+1,m})+\right. \right. \nonumber \\
&+&\left. {\frac{S}{2}}(a_{n,m}^{\dagger }+a_{n,m})(a_{n+1,m}^{\dagger
}+a_{n+1,m})\right) -{\frac{S}{2}}(a_{n,m}^{\dagger
}-a_{n,m})(a_{n+1,m}^{\dagger }-a_{n+1,m})- \nonumber \\ &-&\cos (\theta
_{n,m+1}-\theta _{n,m})\left( (S-a_{n,m}^{\dagger
}a_{n,m})(S-a_{n,m+1}^{\dagger }a_{n,m+1})+\right. \nonumber \\
&+&\left. {\frac{S}{2}}(a_{n,m}^{\dagger }+a_{n,m})(a_{n,m+1}^{\dagger
}+a_{n,m+1})\right) - \nonumber \\ &-&{\frac{S}{2}}(a_{n,m}^{\dagger
}-a_{n,m})(a_{n,m+1}^{\dagger }-a_{n,m+1})-
\nonumber \\
&-&{\frac{\alpha ^{2}}{2}}\left( {\frac{S}{2}}(a_{n,m}^{\dagger
}-a_{n,m})(a_{n+1,m}^{\dagger }-a_{n+1,m})+\right. \nonumber \\
&+&{\frac{S}{2}}(a_{n,m}^{\dagger }+a_{n,m})(a_{n,m+1}^{\dagger
}+a_{n,m+1})\sin \theta _{n,m}\sin \theta _{n,m+1}+ \nonumber \\ &+&\left.
\left. (S-a_{n,m}^{\dagger }a_{n,m})(S-a_{n,m+1}^{\dagger }a_{n,m+1})\cos
\theta _{n,m}\cos \theta _{n,m+1}\right) \right] +\text{ linear terms}
\label{master_sw}
\end{eqnarray}

If all angles $\theta _{n,m}$ in the above expression are given by the
solution of the sin-Gordon equation determining the ground state, the
linear terms will vanish: they represent a static uncompensated force
acting on the spins and must not be present in an equilibrium spin
configuration. In general, Eq.~\ref{master_sw} can not be
diagonalized analytically. Fortunately, we are dealing with a rather
weakly distorted structure and can safely restrict ourselves to
calculating the effect of the KSEA term to the first order in $\delta$.
It is easy to show that the easy $x-y$ plane anisotropy of strength
$\delta$ deforms the spiral in such a way that is $\theta_{n,m}= q n +
(\delta/ 4 \alpha^2) \sin 2qn+O(\delta^2)$, where
$q=\alpha-O(\delta^2)$. Therefore, within our accuracy one can assume
$\theta_{n,m} \simeq
\alpha n$ in Eq. (\ref{master_sw}). In particular the anisotropy
dependence of $\cos (\theta_{n+1,m}-\theta_{n,m}-\alpha) \simeq
\cos (q-\alpha+(\delta/ 4J \alpha^2) \sin 2qn+O(\delta^2))
\simeq 1+O(\delta^2)$ can be disregarded.
With these simplifications and after Fourier transformation
Eq. (\ref{master_sw}) becomes
\begin{eqnarray}
{\cal H}^{(1)}+{\cal H}^{(3)} &=&J\sum_{k_{x},k_{y}}\left[ \left( 4+\delta
-{\frac{\delta }{2}}
\cos k_{y}\right) a(k_{x},k_{y})^{\dagger }a(k_{x},k_{y})-\right.  \nonumber
\\
&-&\left( 2\cos k_{x}+2\cos k_{y}+{\frac{\delta }{2}}\cos k_{y}\right) {
\frac{a(-k_{x},-k_{y})a(k_{x},k_{y})+a(-k_{x},-k_{y})^{\dagger
}a(k_{x},k_{y})^{\dagger }}{2}}+  \nonumber \\
&+&\left( {\frac{\delta }{2}}+{\frac{\delta }{4}}\cos k_{y}\right) \left(
a(k_{x}+2\alpha ,k_{y})^{\dagger }a(k_{x},k_{y})+a(k_{x}-2\alpha
,k_{y})^{\dagger }a(k_{x},k_{y})\right) +  \nonumber \\
&+&{\frac{\delta }{4}}\cos k_{y}\left( {\frac{a(-k_{x}+2\alpha
,k_{y})a(k_{x},k_{y})+a(-k_{x}+2\alpha ,k_{y})^{\dagger
}a(k_{x},k_{y})^{\dagger }}{2}}+\right.  \nonumber \\
&+&\left. \left. {\frac{a(-k_{x}-2\alpha
,k_{y})a(k_{x},k_{y})+a(-k_{x}-2\alpha ,k_{y})^{\dagger
}a(k_{x},k_{y})^{\dagger }}{2}}\right) \right]  \label{sw_lin}
\end{eqnarray}
From this equation we can already qualitatively understand the role of
KSEA interactions. Their main impact is the introduction of terms that
couple magnons with wave vectors that differ by $2\bbox{q}_0$. This
coupling will have the largest effect when acting on a pair of magnons
of equal energies. The result will be discontinuities in the magnon
branches at certain wave vectors, that for the distorted helix become
new zone-boundaries. This picture is very similar to the formation of a
zone structure and zone-boundary energy gaps in a free electron gas,
subject to a weak periodic external potential. Another consequence of
KSEA interaction is the reduction of the energy gap in the
$\bbox{Q}_{\pi ,\pi }$ branch at $k_x=k_y=\pi$ from $2\sqrt{2}DS$ to
$2DS$. However, contrary to the one-dimensional case this gap does not
become zero, i.e. a new Goldstone excitation does not appear in two
dimensions. This can be derived by looking at the part of the Eq.
(\ref{sw_lin}), which does not involve mixing of branches separated by
$2\bbox{q}_0$, and, therefore, yields to the standard analytical
calculation.

To actually calculate the spin wave spectrum we have to find a
transformation of Bose operators that would diagonalize the
Hamiltonian~(\ref{sw_lin}). This transformation must respect Bose
commutation relations, and for a rather general case of helimagnetic
structures is described in detail in Refs.~\onlinecite{Zhitomirsky96}. It
is essentially a Bogolyubov transformation involving a column vector of
four operators: $\hat{a}(k_{x},k_{y})=(a(k_{x}-
\alpha ,k_{y})^{\dagger },a(-k_{x}+\alpha ,-k_{y}),a(k_{x}+\alpha
,k_{y})^{\dagger },a(-k_{x}-\alpha ,-k_{y}))$. The relevant part of the
Hamiltonian (\ref{sw_lin}) can be written as ${\cal H}=(1/2)\sum_{kx,ky}
\hat{a}(k_{x},k_{y})^{\dagger }\hat{V}\hat{a}(k_{x},k_{y})$
, where the $4\times 4$ matrix $\hat{V}$ is given by
\begin{equation}
\hat{V}=\left(
\begin{array}{cccc}
A(k_{x}-\alpha ,k_{y}) & B(k_{x}-\alpha ,k_{y}) & C(k_{x}-\alpha ,k_{y}) &
D(k_{x}-\alpha ,k_{y}) \\ B(k_{x}-\alpha ,k_{y}) & A(-k_{x}+\alpha ,-k_{y})
& D(k_{x}-\alpha ,k_{y}) & C(-k_{x}+\alpha ,-k_{y}) \\ C(k_{x}+\alpha
,k_{y}) & D(k_{x}+\alpha ,k_{y}) & A(k_{x}+\alpha ,k_{y}) & B(k_{x}+\alpha
,k_{y}) \\ D(k_{x}+\alpha ,k_{y}) & C(-k_{x}-\alpha ,-k_{y}) &
B(k_{x}+\alpha ,k_{y}) & A(-k_{x}-\alpha ,-k_{y})
\end{array}
\right)
\end{equation}
Here $A(k_{x},k_{y})=JS\ [2+3\delta -(\delta /2)\cos k_{y}]$, $
B(k_{x},k_{y})=-JS\ [(2+2\delta )\cos k_{x}+2\cos k_{y}+\delta /2\cos
k_{y}]$ , $C(k_{x},k_{y})=JS\ [\delta /2+(\delta /4)\cos k_{y}]$, and $
D(k_{x},k_{y})=JS\ (\delta /4)\cos k_{y}$. To diagonalize the spin wave
Hamiltonian and at the same time ensure the conservation of commutation
relations we have to find a matrix $\hat{Q}$ such that $\hat{Q}^{\dagger
}\hat{V}\hat{Q}$ is diagonal, while $\hat{Q}^{\dagger
}\hat{g}\hat{Q}=\hat{g}$, where $\hat{g} $ is the diagonal matrix with
diagonal elements $(1,-1,1,-1)$. This is equivalent to diagonalizing the
matrix $\hat{g}\hat{V}$. \cite{Zhitomirsky96}

To obtain numerical results that could be directly compared to our
measurements on Ba$_{2}$CuGe$_{2}$O$_{7}$ we used the independently
measured values for $J=0.96$~meV and $D/J\approx \arctan(D/J)\equiv
\alpha=\frac{32}{31}\phi=0.177$. A numerical diagonalization of
$\hat{g}\hat{V}(0,0)$ was performed using Mathematica software package
to yield the eigenvalues ${\cal E}_{0}=0.172$~meV, ${\cal E}_{1}=0.297$
meV and ${\cal E}_{2}=0.171$ meV. These are the energies of the
$\bbox{Q}_{\pi ,\pi }$ (${\cal E}_{0}$), and $\bbox{Q}_{\pi ,\pi
}\pm\bbox{q}_{0}$ (${\cal E}_{1}$, ${\cal E}_{2}$) branches at the AFM
zone center $\bbox{Q}_{\pi ,\pi }$. The splitting was predicted to be
$2\delta_{\pi ,\pi }=({\cal E}_{1}-{\cal E}_{2})=0.12$ meV. This value
is indistinguishable from the actual splitting observed in
Ba$_{2}$CuGe$_{2}$O$_{7}$, quoted in the previous section. We can also
calculate the splitting in the $\bbox{Q}_{\pi ,\pi }$ branch at
$\bbox{Q}_{\pi ,\pi }\pm \bbox{q}$: $2\delta _{\bbox{q}_{0}}=0.049$ meV.
Experimentally, this splitting was not observed in zero field, but is
small enough to be well within the experimental error bars. At higher
fields the discontinuity at this wave vector becomes apparent (see
Section~(\ref{incom})). The gap in the $\bbox{Q}_{\pi ,\pi }$-branch,
$\Delta_{\pi ,\pi }={\cal E}_{0}=0.172$~meV, is also in very good
agreement with the INS measurements. Entire dispersion branches
calculated numerically using the technique described above are shown in
Fig.~10(b). They can be also seen as solid lines in Fig.~\ref{disp0} and
apparently are in very good agreement with experimental data.

\subsubsection{Spin-wave spectrum in the spin-flop phase}
As we have already mentioned, we presently do not have theoretical
results for the spin-wave dispersion in the presence of an external
magnetic field. However, we can make some predictions for the spin-wave
spectrum in the spin-flop phase (i.e., for $H>H_{c}$). After some
tedious calculations that are omitted here, but are very similar to
those performed in Section~\ref{DMY}, one arrives at the result that the
contribution of the Dzyaloshinskii-Moriya term for the $x$ ($y$) bonds
is proportional to $\sin k_{x}$ ($\sin k_{y}$) and therefore {\it
exactly} vanishes at the AFM zone-center. In order to calculate the
additional energy gap $\Delta_{{\rm c}}$ in the spin-flop phase we thus
need to consider only the KSEA terms. At $\bbox{Q}=\bbox{Q}_{\pi ,\pi }$
(long-wavelength limit) the effect of KSEA interactions is identical to
that of conventional easy-plane exchange anisotropy. The spectrum of a
Heisenberg AFM with such anisotropy in a magnetic field is
well-known.\cite{Akhiezer} Both field and anisotropy split the two-fold
degenerate magnons in a Heisenberg system to give a gapless mode with
linear dispersion and an ``optical'' mode with the energy gap at
$\bbox{Q}_{\pi ,\pi }$ given by:
\bigskip
\begin{eqnarray*}
{\hbar \omega }(\bbox{Q}_{\pi ,\pi }) &=&
\sqrt{\Delta _{{\rm c} }^{2}+(g\mu_{B}H)^{2}}\text{,} \\
\Delta _{c} &=&2\sqrt{2} JS \sqrt{2\delta}=2\sqrt{2} DS \text{.}
\end{eqnarray*}
Substituting the known numerical values into this formula we obtain $\Delta
_{c}=0.24$ meV, which is in reasonable agreement with the experimental
result $\Delta _{{\rm c}}=0.28$ meV.

\section{Discussion}

We see that both the static and dynamic properties of
Ba$_{2}$CuGe$_{2}$O$_{7}$ are quantitatively consistent with the
presence of KSEA interactions. To be more precise, the experimental data
unambiguously indicate the presence of an easy-plane anisotropy of
exactly the same strength as predicted by the KSEA mechanism. It is
important to stress that in a {\it slowly} rotating helix it is
impossible to distinguish experimentally between single-ion easy-plane
anisotropy of type $
\sum_{n,m}\left( S_{n,m}^{z}\right) ^{2}/2$, two-ion anisotropy ($\sum_{n,m}
\left[ S_{n,m}^{z}S_{n+1,m}^{z}+S_{n,m}^{z}S_{n,m+1}^{z}\right]/2$ ) or
KSEA-type anisotropy ( $\sum_{n,m}\left[
S_{n,m}^{y}S_{n+1,m}^{y}+S_{n,m}^{x}S_{n,m+1}^{x}\right] $ ) of the same
strength. Indeed, the difference between a pair of easy axes (KSEA term)
and an easy plane (conventional single-ion or two-ion anisotropy)
becomes apparent only when the period of the structure is comparable to
the nearest-neighbor spin-spin separation, i.e., is only manifested in
lattice effects. Alternatively, it can be observed in strong magnetic
fields when the canting of spins towards the field direction becomes
substantial. In Ba$_{2}$CuGe$
_{2}$O$_{7}$ , where the magnetic structure has a rather long periodicity,
and where even at the critical field the uniform magnetization (spin
canting) is small, these effects are expected to be insignificant. It is
entirely possible that the weak ``quadrupolar'' in-plane anisotropy seen in
horizontal-field experiments\cite{ZM97BACUGEO-B} is in fact such a lattice
effect. Note that its strength is extremely small, of the order of $7
\times 10^{-9}$~eV,
and yet it can be reliably measured in a diffraction experiment where a
magnetic field is applied in the $(ab)$ crystallographic plane at different
angles to the $a$ axis \cite{ZM97BACUGEO-B}. Another possible manifestation
of lattice effects is the intermediate phase seen just before the CI
transition in a magnetic field applied along the $c$ crystallographic axis
\cite{ZM98BACUGEO}. A further study of these phenomena that distinguish
between KSEA and other type of anisotropy in Ba$_{2}$CuGe$_{2}$O$_{7}$ is
an interesting topic for future experimental and theoretical work.

One final comment has to be made in reference to another possible player in
the spin Hamiltonian for Ba$_{2}$CuGe$_{2}$O$_{7}$: dipolar interactions.
In principle, the ever-present dipolar term can and will influence both the
ground state spin configuration and the spin wave spectrum in Ba$_{2}$CuGe$
_{2}$O$_{7}$. Its effect however is expected to be insignificant compared
even to the weakest terms that we have considered in our treatment. Indeed,
nearest-neighbor spins in Ba$_{2}$CuGe$_{2}$O$_{7}$ are separated by
$\Lambda=a/
\sqrt{2}\approx 6$ \AA . For two nearest-neighbor spins the energy of
dipolar coupling is of the order of $\left( g\mu _{B}\right)
^{2}/\Lambda^{3}\approx 1\,\mu$eV. \ This is an order of magnitude less than the
smallest energy scale in our model, which is the strength of KSEA
anisotropy $J\alpha ^{2}/2\approx 15\,\mu$eV. Moreover, in an
almost-antiferromagnetic structure long-range dipolar interactions will be
heavily suppressed by the sign-alternation in the contribution of
individual pairs of interacting spins. Our neglecting dipolar interactions
in all the derivations above is thus justified.

In summary we have demonstrated that KSEA interactions can result in very
interesting measurable effects, and that no {\it additional} anisotropy is
needed to reproduce the behavior observed in Ba$_{2}$CuGe$_{2}$O$_{7}$.

\acknowledgements
This study was supported in part by NEDO (New Energy and Industrial
Technology Development Organization) International Joint Research Grant and
the U.S. -Japan Cooperative Program on Neutron Scattering. Work at BNL was
carried out under Contract No. DE-AC02-98CH10886, Division of Material
Science, U.S. Department of Energy. Experiments at NIST were partially
supported by the NSF under contract No. DMR-9413101.

\begin{figure}
\caption{Typical elastic scans along the $(1,1,0)$ direction in the vicinity of
the antiferromagnetic zone-center $(1,0,0)$, measured in
\bacugeo at $T=0.35$~K in zero field (top) and in a $H=1.6$~T magnetic field applied
along the $(0,0,1)$ direction (bottom). Note the logarithmic scale on
the $y$-axis. The solid lines are guides for the eye. The arrows show
the positions of the principal magnetic Bragg reflections at
$(1+\zeta,\zeta,0)$, characteristic of a spiral spin structure, and the
3rd harmonic at $(1+3\zeta,3\zeta,0)$, a signature of a slight
distortion of the helicoid. Insert: a schematic of the magentic
structure showing a single Cu-plane in \bacugeo.}
\label{diff}
\end{figure}

\begin{figure}
\caption{Transverse elastic scans through the 1st-order (a) and 3rd-order(b)
magnetic Bragg reflections measured in \bacugeo at $T=1.3$~K and
$H=1.6$~T. Solid lines are Gaussian fits to the data. The shaded
Gaussian represents the calculated experimental $Q$-resolution. The
intrinsic angular width of both peaks is $\approx 20^{\circ}$ as seen
from the $(1,0,0)$ antiferromagnetic zone-center.}
\label{diff2}
\end{figure}

\begin{figure}
\caption{(b) Measured field dependence of the magnetic Bragg peak intensities in \bacugeo at
$T=0.35$~K. Solid and open circles show the beha
vior of the 1st-order and
3rd-order incommensurate Bragg peak integrated intensities, respectively.
The intensity of the commensurate peak at the antiferromagnetic zone-center
is plotted in open triangles. (b) Square root of the ratio of intensities
of the 3rd and 1st harmonic plotted against the square of the applied
magnetic field. The solid and dashed lines show the theoretical prediction
for the DM-only and DM+KSEA models (c) Measured intensity of the 3rd
harmonic plotted against the normalized incommensurability parameter
$\zeta$. The solid line shows the prediction of the DM+KSEA model.}
\label{diffresult}
\end{figure}

\begin{figure}
\caption{ Typical inelastic scans measured in \bacugeo in the two
experimental runs, at ILL (top) and NIST (bottom), respectively. The
heavy solid line is a multiple-Gaussian fit to the data, and the shaded
curves represent the individual Gaussians. The gray area in the top
panel shows the position of a ``Bragg-tail'' spurious peak.}
\label{exdata}
\end{figure}

\begin{figure}
\caption{ Spin wave dispersion curves measured in
\bacugeo in zero magnetic field.
The data collected at $T=0.35$~K and $T=1.5$~K are combined in this plot.
The solid lines are parameter-free theoretical curves as described in the
text. Dashed lines are guides for the eye and the solid circles on the
abscissa show the positions of the observed magnetic Bragg peaks.}
\label{disp0}
\end{figure}

\begin{figure}
\caption{ Spin wave dispersion curves measured in \bacugeo in a $H=1$~T magnetic
field applied along the $(001)$ direction at $T=0.35$~K. Solid lines are
guides for the eye. Dashed lines are as in Fig.~\protect{\ref{disp0}}.}
\label{disp1}
\end{figure}

\begin{figure}
\caption{ Spin wave dispersion curves measured in
\bacugeo in a $H=1.5$~T magnetic
field applied along the $(001)$ direction at $T=0.35$~K. The lines and
symbols as in previous figures. Note the additional
branch in the spectrum.}
\label{disp15}
\end{figure}

\begin{figure}
\caption{ Spin wave dispersion curves measured in \bacugeo in
a $H=2.5$~T magnetic field applied along the $(001)$ direction at
$T=0.35$~K (commensurate spin-flop phase). The upper solid line is a fit
to Eq.~\protect{\ref{edisp4}}. The lower line is a linear fit.}
\label{disp25}
\end{figure}

\begin{figure}
\caption{
Field dependence of the incommensurability parameter $\zeta$, as
previously measured in Ba$_2$CuGe$_2$O$_7$. The solid curve is plotted
using Eqs.~\protect{\ref{zeta1}},\protect{\ref{zeta2}}, that takes into
account KSEA anisotropy. The dashed curve is the prediction of the
DM-only model.}
\label{maslov1}
\end{figure}

\begin{figure}
\caption{
Theoretical predictions for the spin wave dispersion along the
$x$-axis in (a) the DM-only model, (b) DM+KSEA model.
The effect of the KSEA term is to couple magnons separated by
$2{\bf \it q}_0$, which leads to the appearance of new gaps in the spectrum,
and reduce the gap in ${\bf \it Q}_{\pi,\pi}$ branch.}
\label{tfig1}
\end{figure}

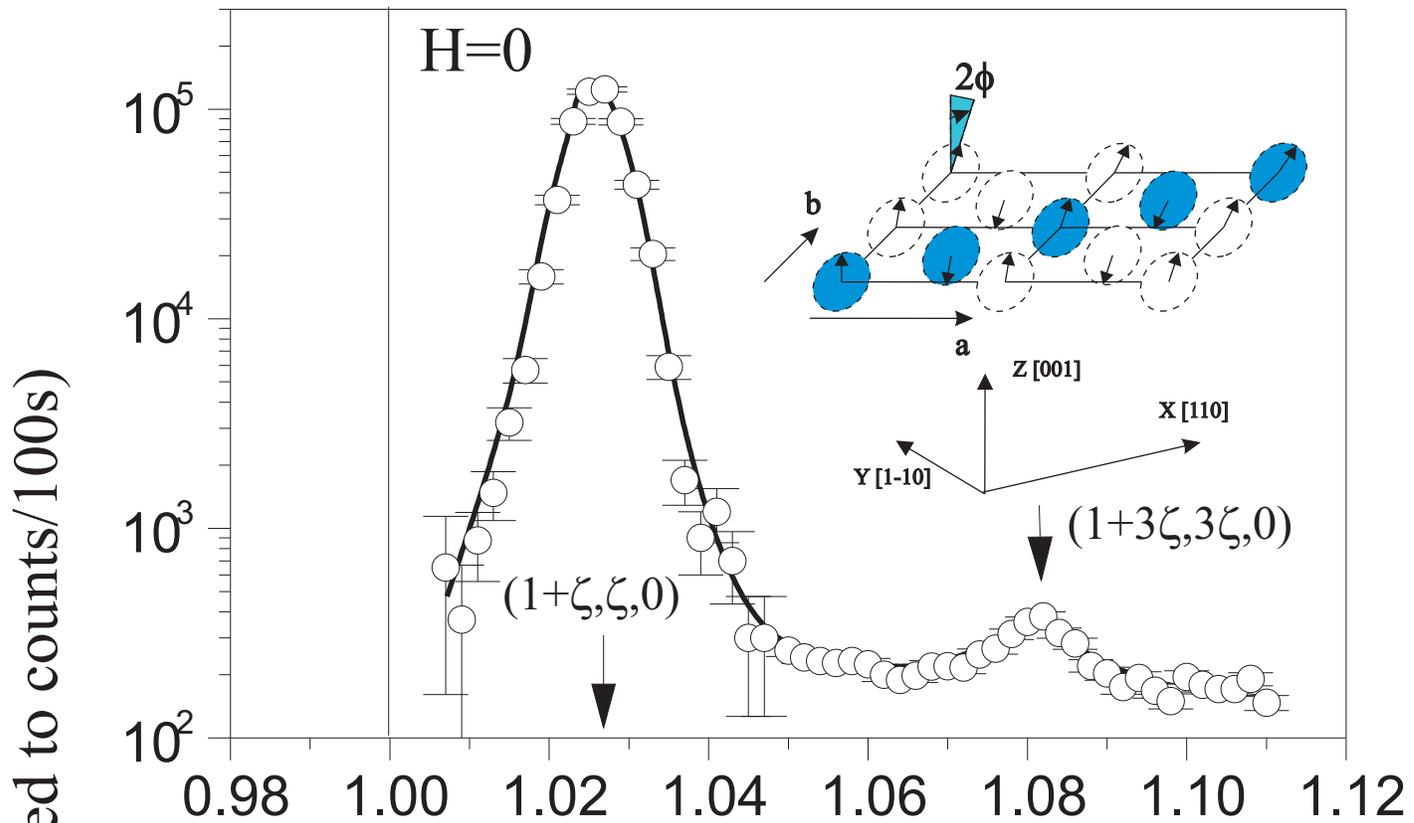
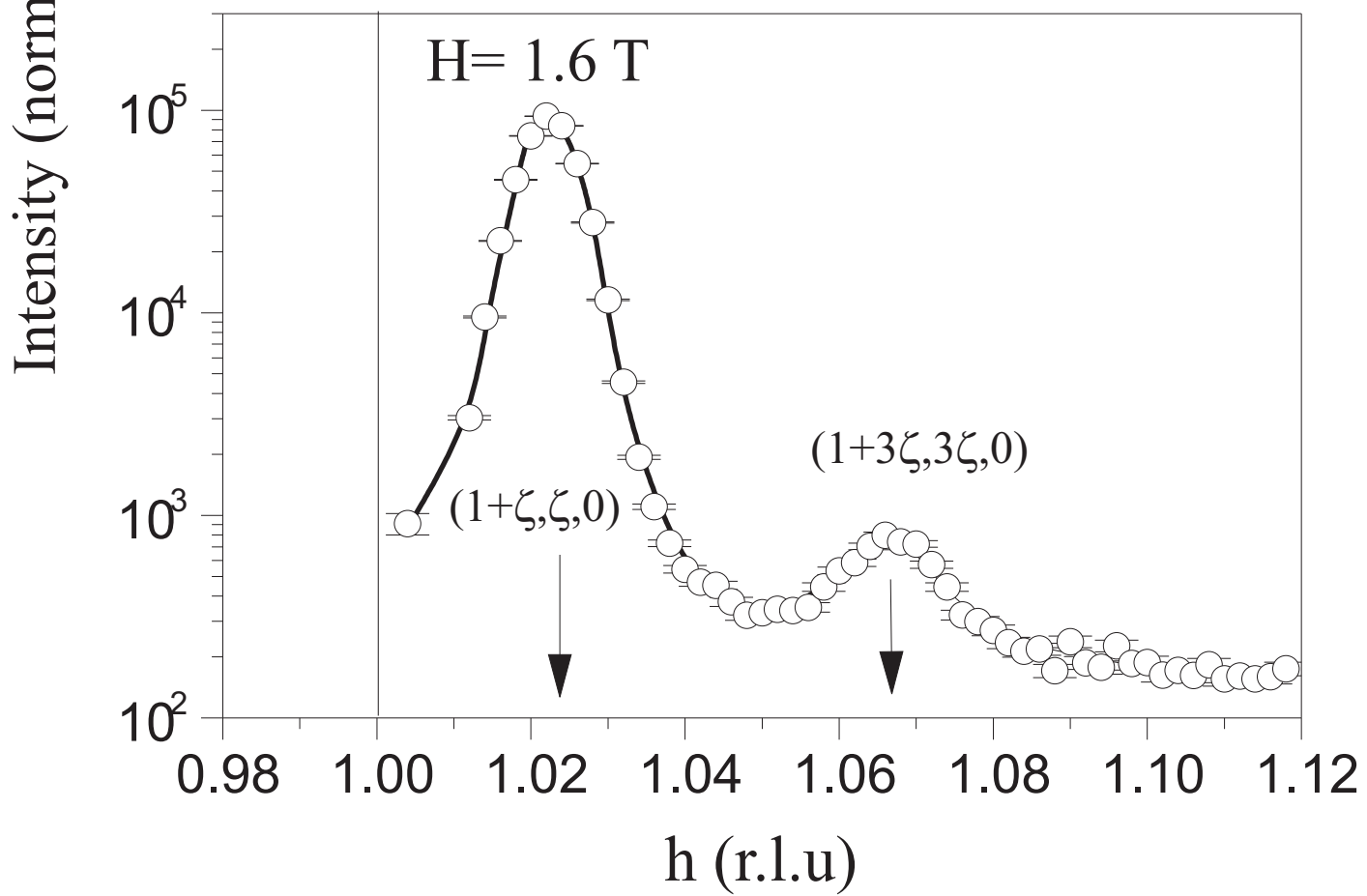
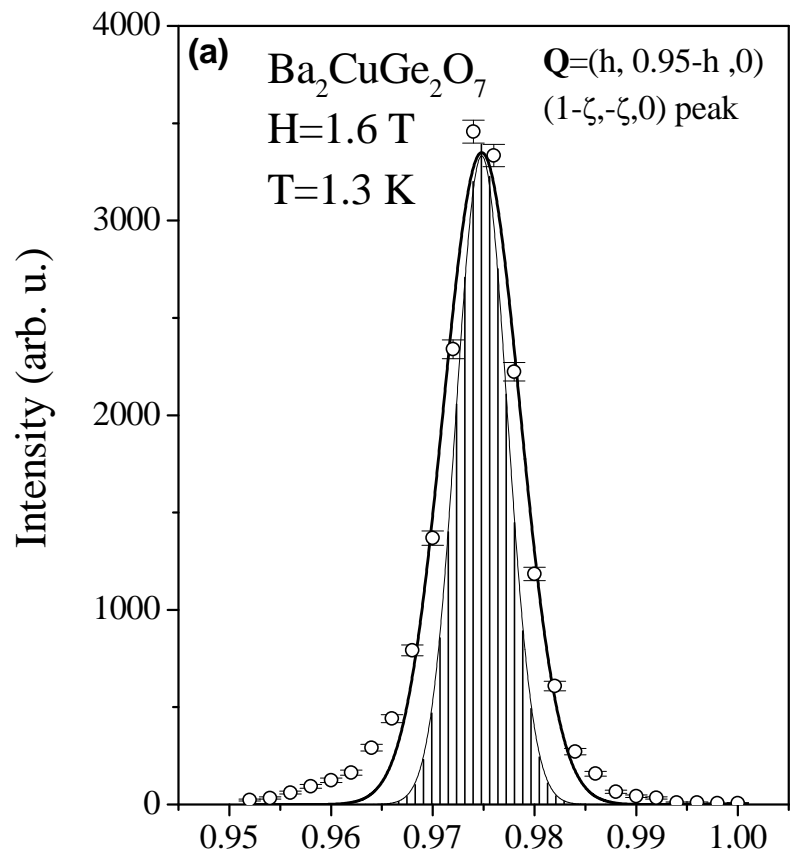
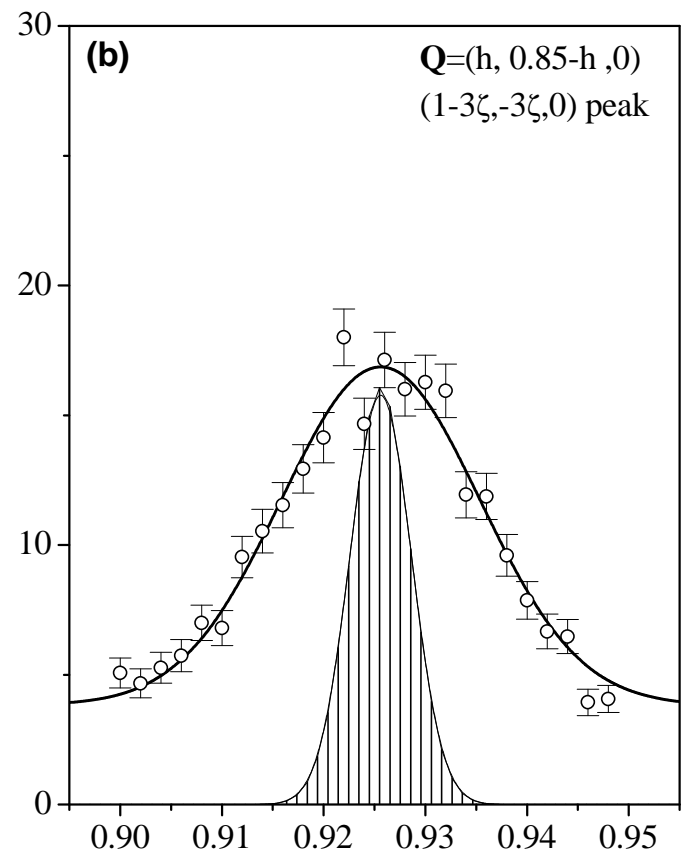
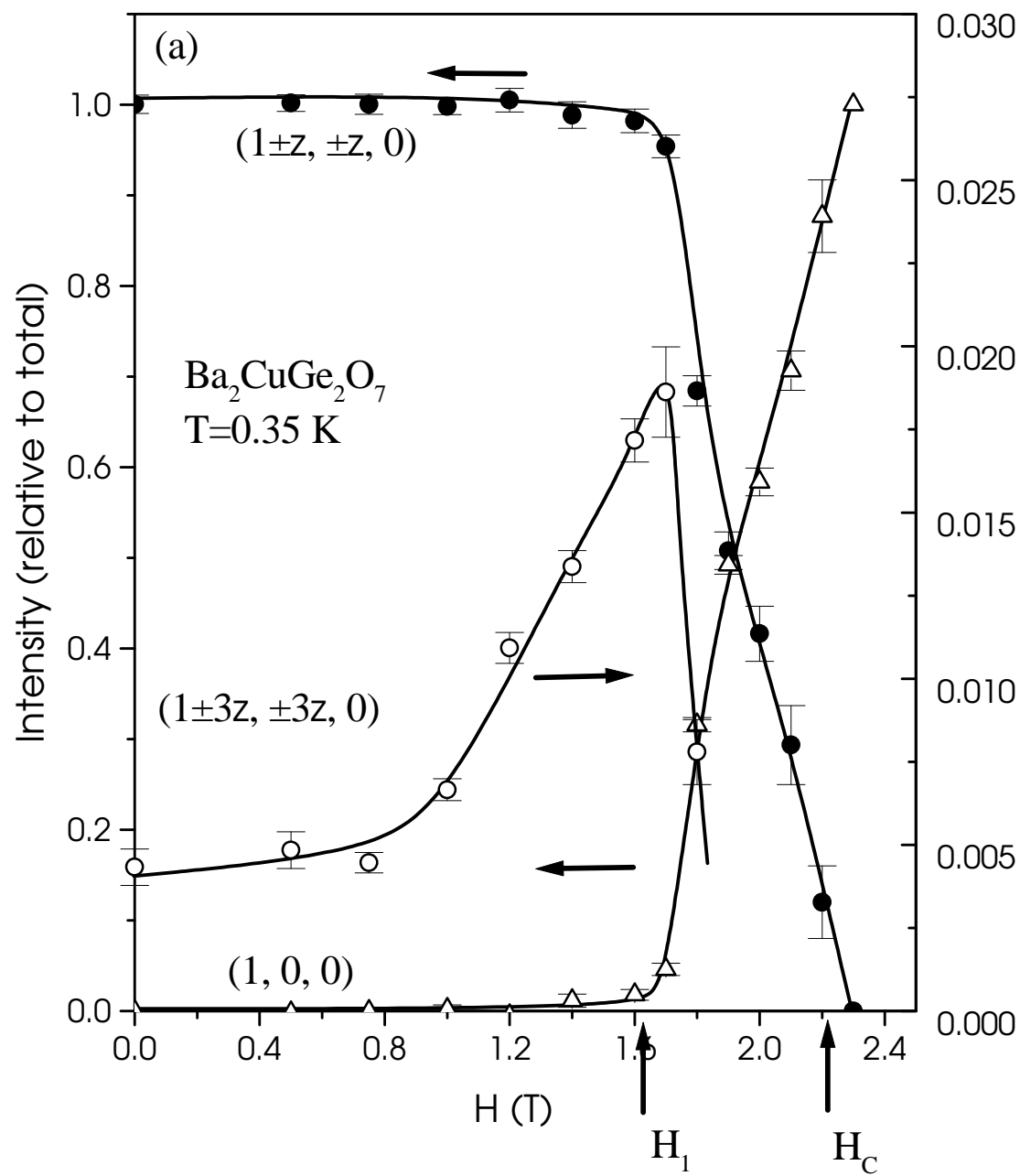
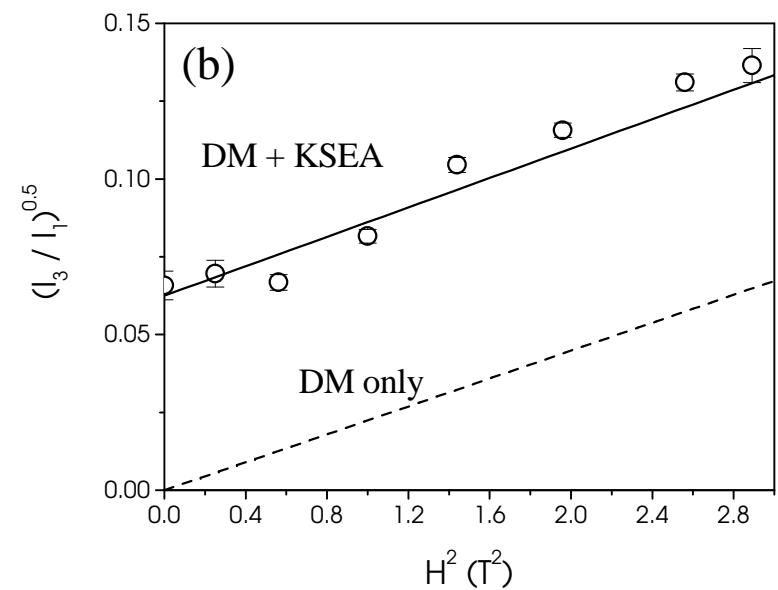
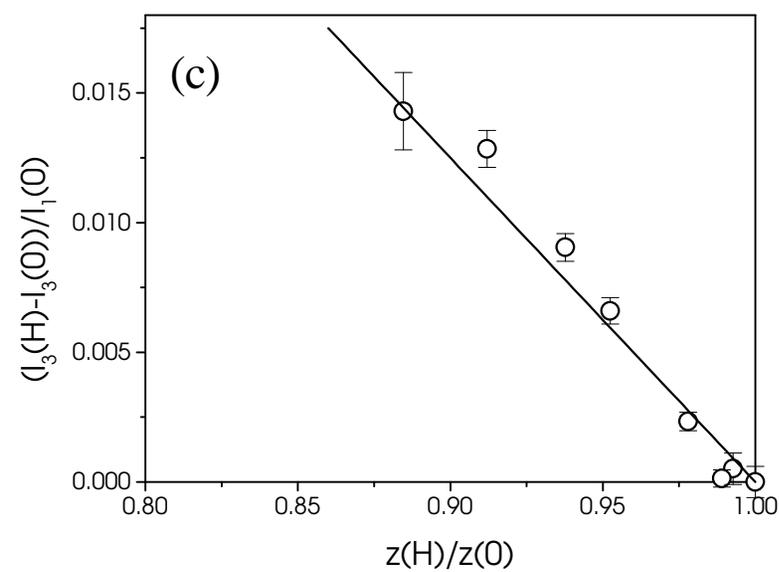
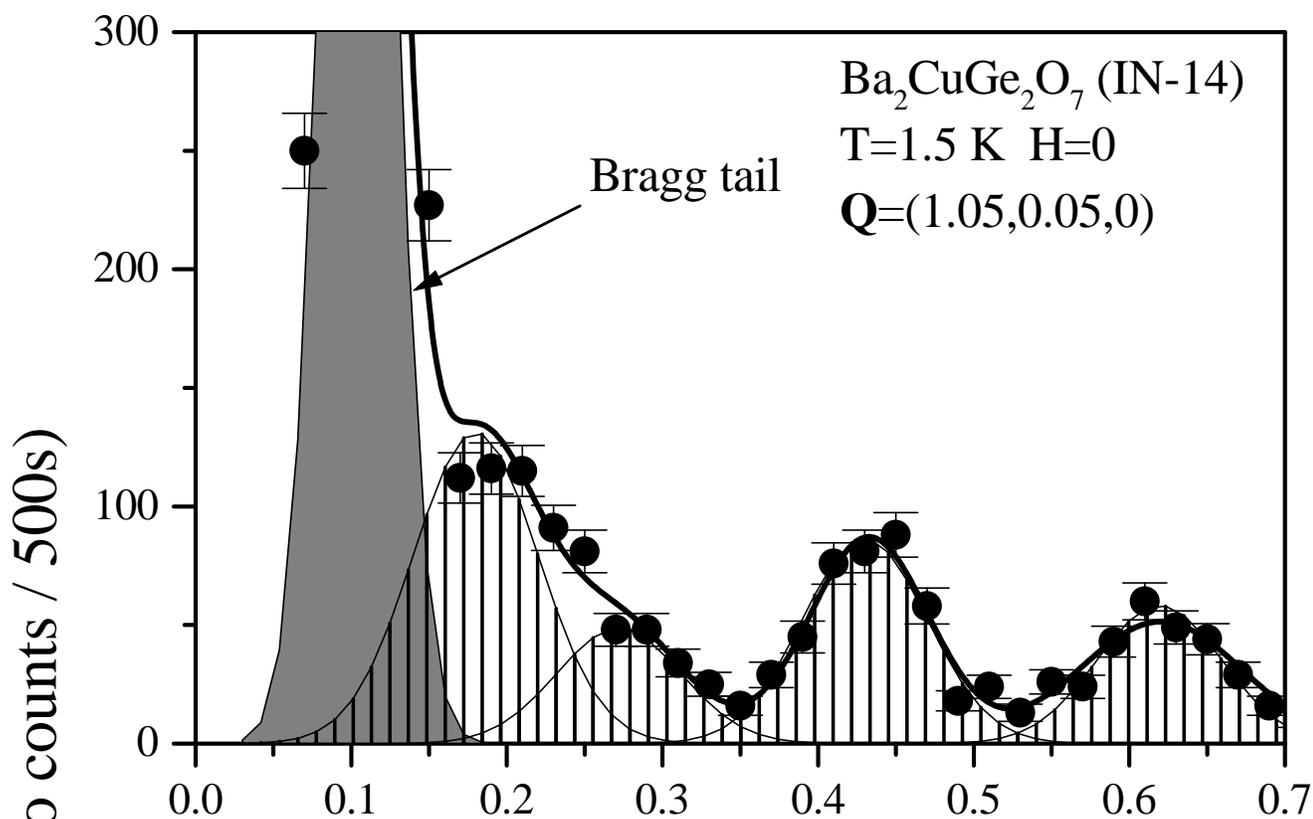
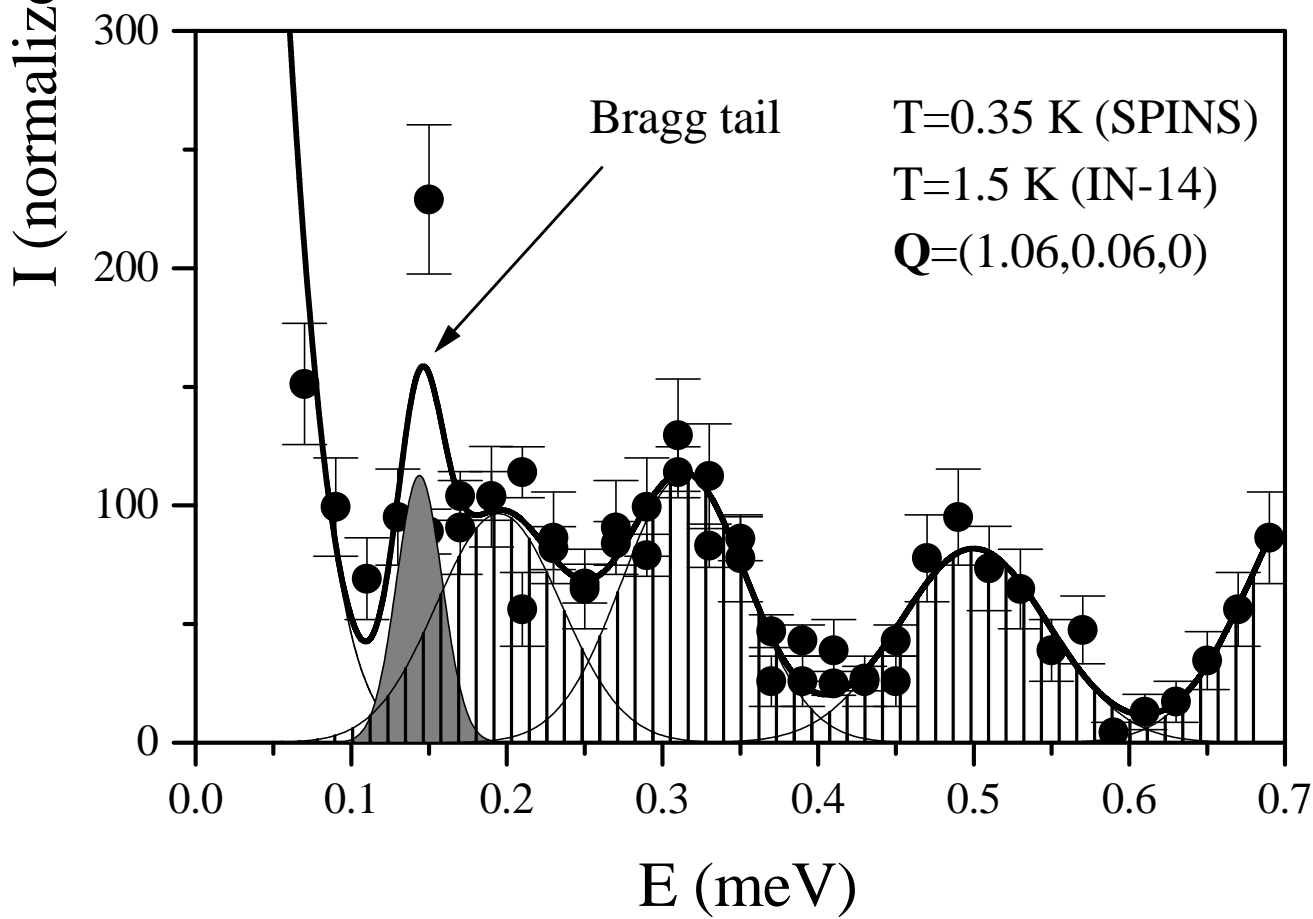
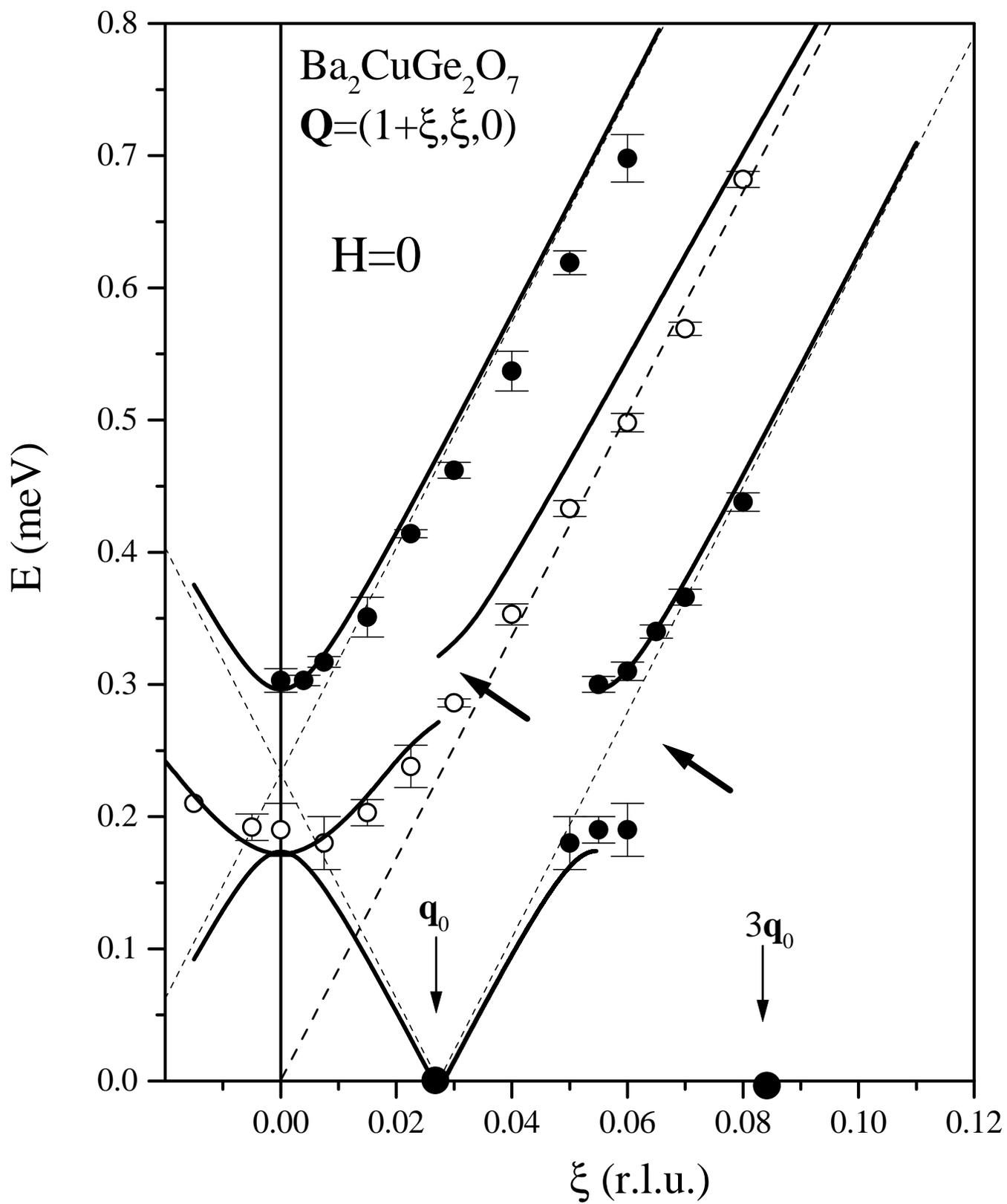
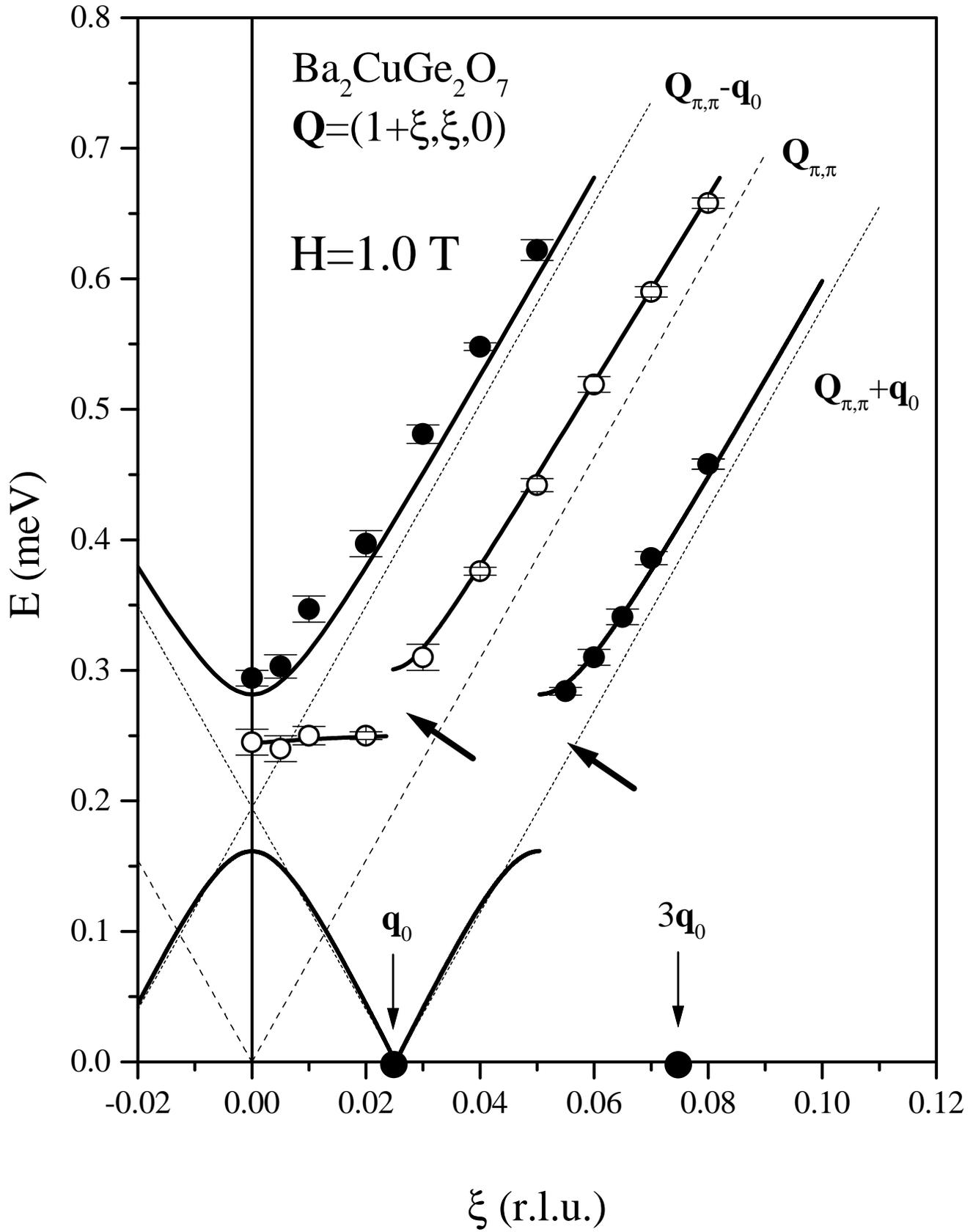
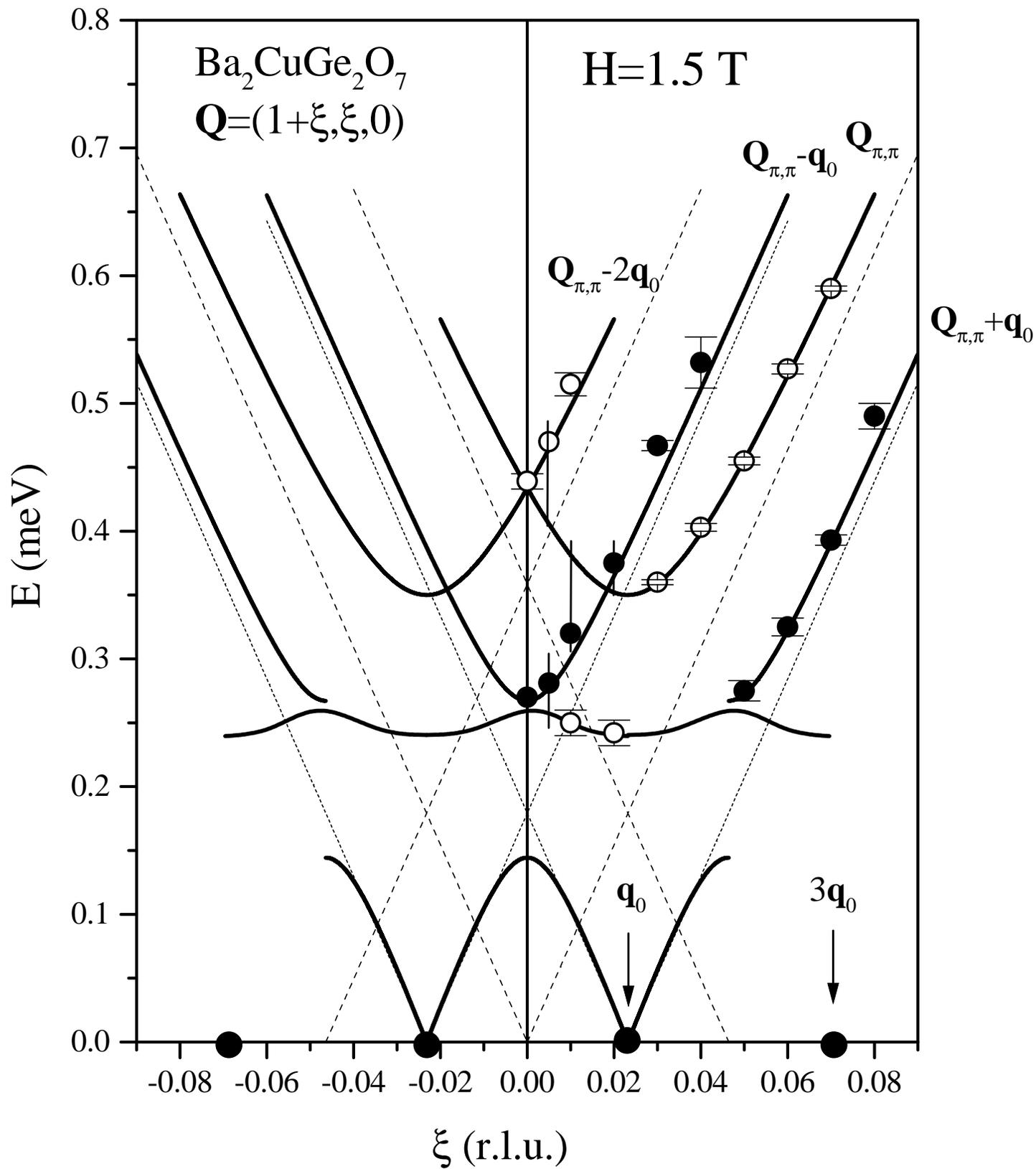
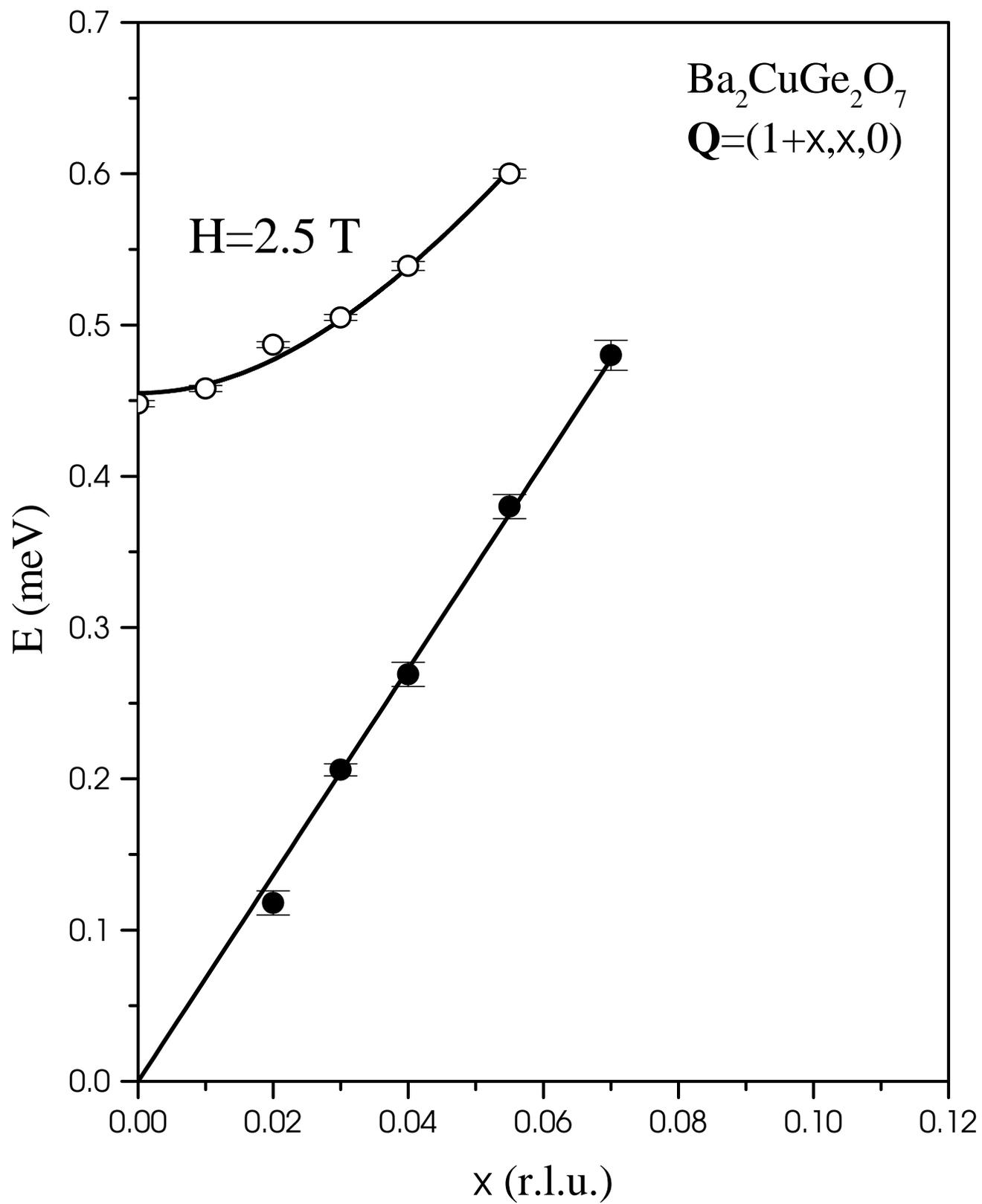
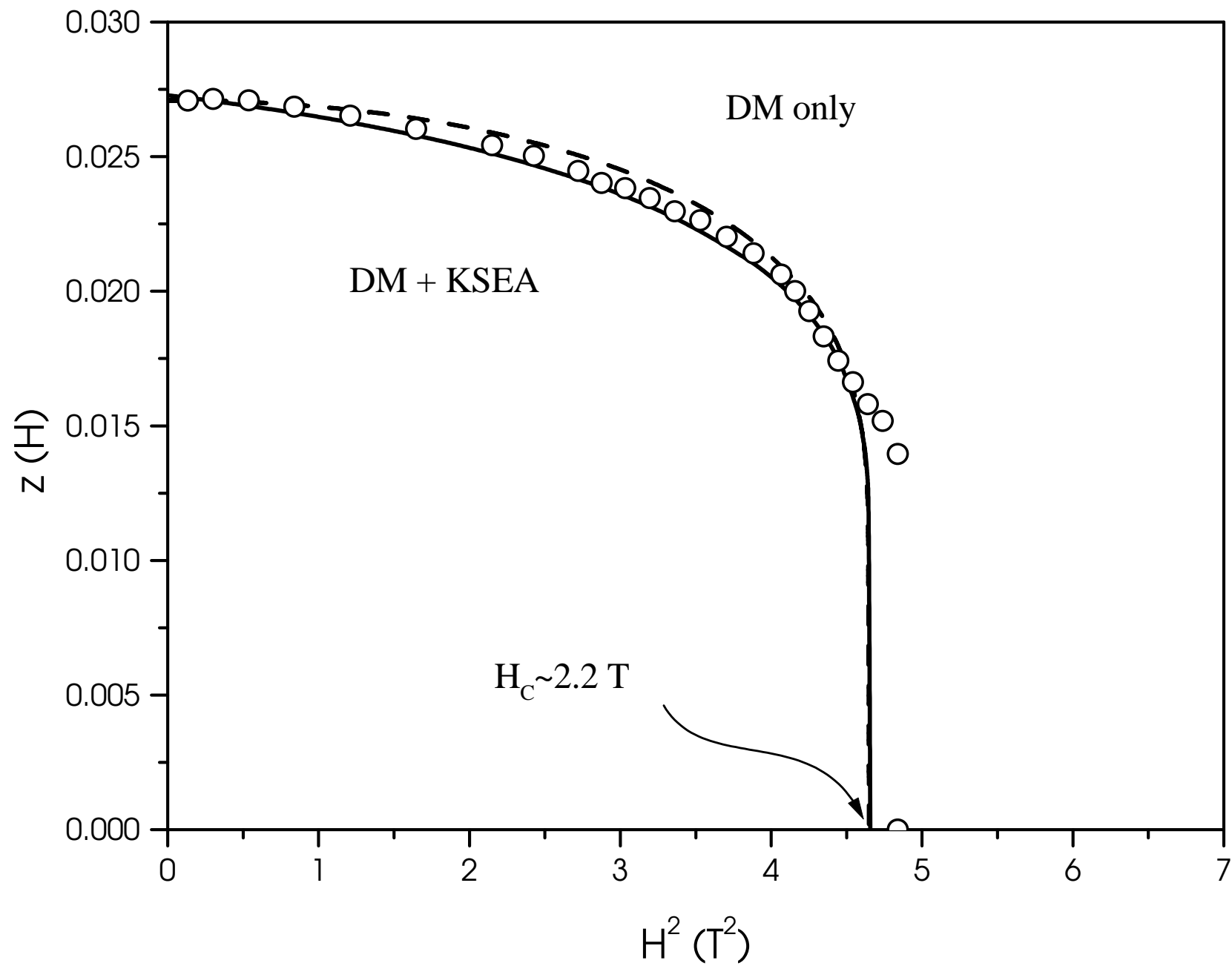
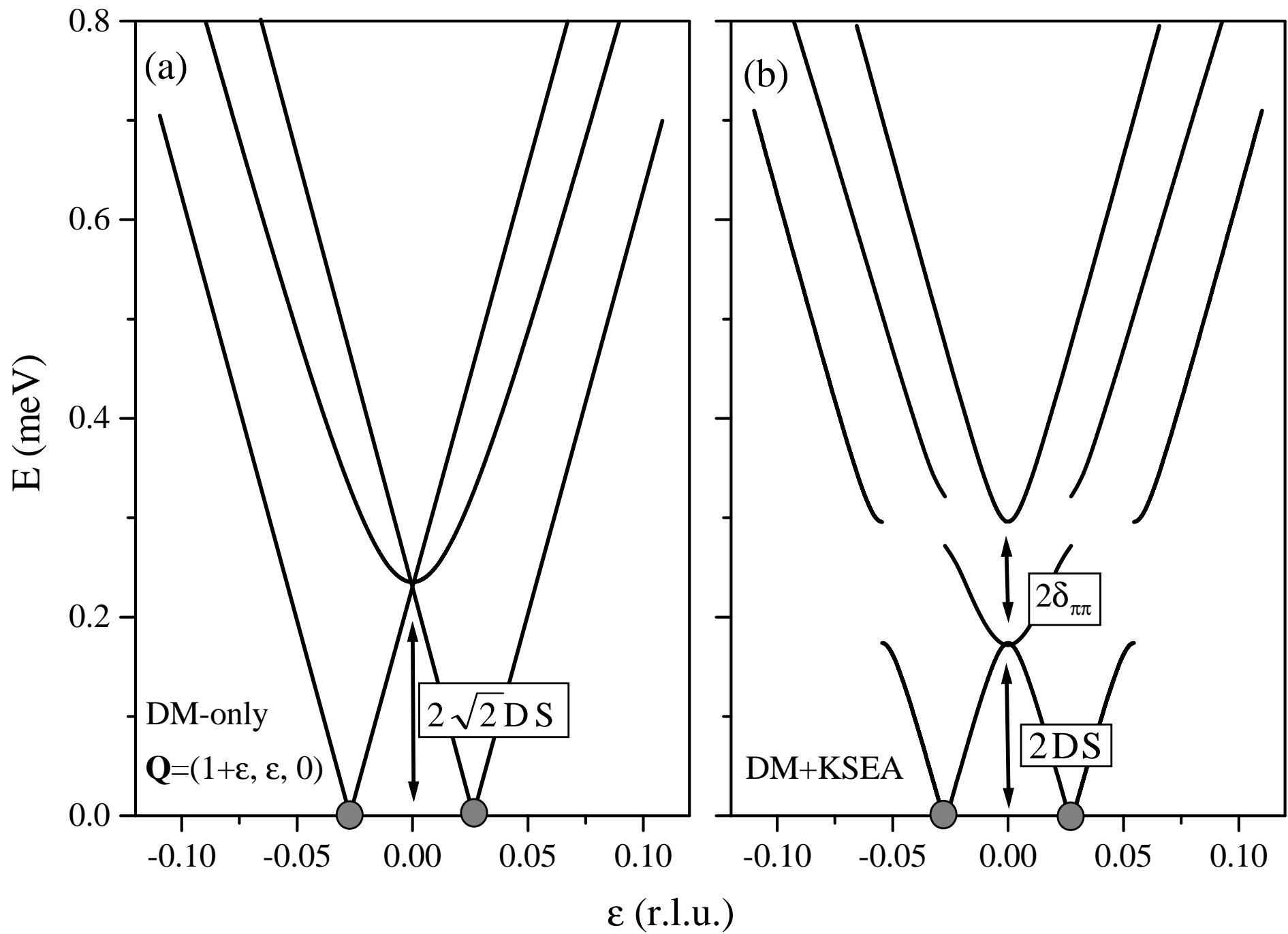
\end{document}